\documentclass[a4paper,11pt]{article}
\pdfoutput=1 

\usepackage{jcappub} 

\usepackage[T1]{fontenc} 

\usepackage{amssymb,amsfonts,amsmath,paralist,xspace} %
\usepackage{units}

 \newcommand{\be}
{\begin{equation}} \newcommand{\ee} {\end{equation}}   
\newcommand{\eq}[1]{\begin{equation} #1 \end{equation}}
\def\be{\begin{eqnarray}} \def\ee{\end{eqnarray}} 
\newcommand{\dm}{{\textsc{dm}}} 

\newcommand{\fov}{\mathrm{fov}} 
 
\title{\boldmath Influence of $\sim$7~keV sterile neutrino dark matter on the process of reionization}

\author[a,b]{Anton~Rudakovskyi,}
\author[c,a,1]{Dmytro~Iakubovskyi\note{Corresponding author.}}
\affiliation[a]{Bogolyubov Institute of Theoretical Physics, Metrologichna Str. 14-b, 03680, Kyiv, Ukraine}
\affiliation[b]{Taras Shevchenko National University of Kyiv, Physics Department, Glushkova ave. 2, Kyiv, Ukraine}
\affiliation[c]{Discovery Center, Niels Bohr Institute, Blegdamsvej 17, Copenhagen, Denmark}

\emailAdd{iakubovskyi@nbi.ku.dk}

\abstract{Recent reports of a weak unidentified emission line at $\sim$3.5~keV found in spectra of several 
matter-dominated objects may give a clue to resolve the long-standing problem of dark matter. 
One of the best physically motivated particle candidate able to produce such an extra line 
is sterile neutrino with the mass of $\sim$7~keV. Previous 
works show that sterile neutrino dark matter with 
parameters consistent with the new line measurement modestly affects structure formation compared to conventional 
cold dark matter scenario.
In this work, we concentrate for the first time on contribution of the sterile neutrino dark matter 
able to produce the observed line at $\sim$3.5~keV, to the process of reionization.
By incorporating dark matter power spectra for $\sim$7~keV sterile 
neutrinos
into extended semi-analytical `bubble' model of reionization
we obtain that such sterile neutrino dark matter would produce 
significantly sharper reionization
compared to widely used cold dark matter models,
impossible to `imitate' within the cold dark matter scenario under any reasonable choice of our 
model parameters, and would have a clear tendency of lowering both the redshift 
of reionization and the electron scattering optical depth (although the difference is 
still below the existing model uncertainties).
Further dedicated studies of reionization (such as 21~cm measurements or studies of kinetic 
Sunyaev-Zeldovich effect) will thus be essential for reconstruction of particle candidate 
responsible the $\sim$3.5~keV line.}

\begin{document}
\maketitle
\flushbottom

\section{Introduction}
\label{sec:intro}

The nature of missing mass of the Universe -- presumably in form of gravitating \emph{dark matter} -- 
is largely unknown. For example, Standard Model of particle physics does not provide a viable candidate 
for a dark matter particle. If the dark matter consists of elementary particles,
they may have non-zero initial velocities affecting the structure formation due 
to their `free streaming' from potential wells, see e.g.~\cite{Szalay:76}. 
The most commonly used type of dark matter
is cold dark matter (CDM) where free streaming effects are negligible. 
Although there are well-motivated particle 
physics candidates for CDM such as weakly interacting massive particles 
(WIMPs)~\cite{Lee:77,Goldberg:83,Ellis:84,Servant:02}
or axions\footnote{Note that the spectrum of \emph{ultra-light} axion-like dark matter particles 
significantly differs from that for CDM, see~\cite{Marsh:15} for a recent review.}~\cite{Preskill:83,Abbott:83,Dine:83}, they by no means exhaust the complete list of dark 
matter particle candidates, see e.g. reviews~\cite{Bertone:04,Bergstrom:09,Feng:10,Drees:12,Bertone:16}.

For example, one of the best physically-motivated dark matter candidates is the right-handed (sterile) 
neutrino~\cite{Dodelson:93,Shi:98}, see also an extensive recent review~\cite{Adhikari:16}. 
It has been shown that the minimal extension of the Standard Model by three 
sterile neutrinos (dubbed $\nu$MSM)~\cite{Asaka:05a,Asaka:05b} can be responsible for dark matter 
as well as for a number of another observed phenomena including neutrino oscillations and 
observed asymmetry between matter and anti-matter in the Universe, see~\cite{Boyarsky:09a} for a review.
Formation of dark matter sterile neutrinos in early Universe primarily depends on production mechanism 
because they have never reached thermal equilibrium, contrary to WIMPs or Standard Model (`active') neutrinos. 
In the $\nu$MSM, the lightest sterile neutrinos responsible for dark 
matter are produced by oscillations of `active' neutrinos in primeval plasma with a substantial amount of 
lepton asymmetry generated by earlier decays of two heavier sterile 
neutrinos~\cite{Shaposhnikov:08,Laine:08,Canetti:12a,Canetti:12b}.
As a result, the properties of sterile neutrino dark matter in the $\nu$MSM are determined by three parameters --
the mass of dark matter particle $M_s$, its mixing angle $\theta$ with Standard Model neutrinos
and the value of lepton asymmetry during the particle production.

The parameters $M_s$ and $\theta$ can be reconstructed from the 2-body radiative decay line flux $F_\gamma$ 
at energy $E_\gamma$ expected from direction of dark matter haloes 
located at distance $D_L$ with dark matter mass inside the instrument's field-of-view $m_\dm^\fov$, 
see e.g.~\cite{Boyarsky:07a}:
\eq{
\sin^2(2\theta) = 7.8\times 10^{-10} 
\left(\frac{F_\gamma}{10^{-6}~\unit{ph/cm^2/s}}\right)
\left(\frac{D_L}{1~\unit{kpc}}\right)^2 
\left(\frac{10^6~\unit{M_\odot}}{m_\dm^\fov}\right)
\left(\frac{1~\unit{keV}}{M_s}\right)^4,
\label{eq:m-theta}
}
$M_s = 2E_\gamma$. 
A possible line from decaying dark matter seen at $E_\gamma \simeq 3.55$~keV has been reported 
by~\cite{Bulbul:14a,Boyarsky:14a,Boyarsky:14b,Urban:14,Iakubovskyi:15b}, see also recent 
reviews~\cite{Iakubovskyi:15c,Adhikari:16}.
The observed line parameters are consistent with sterile neutrino parameters
$M_s \simeq 7.1$~keV and $\sin^2(2\theta) \simeq (2-20)\times 10^{-11}$ with the preferred region of 
$\sin^2(2\theta) \simeq (4.5-6)\times 10^{-11}$~\cite{Boyarsky:14b,Iakubovskyi:15b,Ruchayskiy:15}.
The corresponding values of lepton asymmetry obtained under assumption that such sterile neutrino 
constitutes the bulk of dark matter in the Universe are then used to produce the power spectra of 
sterile neutrino dark matter~\cite{Abazajian:14,Venumadhav:15,Lovell:15}. The obtained dark matter 
power spectra and their analogues are then used to investigate 
various outputs of cosmological structure formation such as subhalo number 
counts~\cite{Abazajian:14,Horiuchi:15,Bose:16a,Schneider:16} and subhalo velocity 
function~\cite{Bozek:15,Horiuchi:15,Bose:16a} in the Local Group, central mass density 
distributions in nearby haloes and subhaloes~\cite{Bozek:15,Bose:15,Horiuchi:15,Bose:16a}, 
mass of Milky Way halo hosting observable number of satellite galaxies~\cite{Lovell:15}, 
high-$z$ galaxy counts~\cite{Abazajian:14} and 3D matter power 
spectra~\cite{Bozek:15,Schneider:16}, 1D matter power spectra probed by Ly-$\alpha$ 
observations~\cite{Schneider:16},
halo mass functions~\cite{Bose:15}, concentration-mass scaling relation for dark matter 
haloes~\cite{Schneider:14,Bose:15,Li:15,Ludlow:16}, 
maximal values of dark matter phase-space density in dwarf spheroidal galaxies~\cite{Abazajian:14},
shapes and spins of dark matter haloes~\cite{Bose:15}, mass assembly history and infall time of the Fornax 
dwarf spheroidal galaxy~\cite{Wang:15},
anomalous flux ratios in strong lensed systems~\cite{Kamada:16}, weak lensing maps of galaxy 
clusters~\cite{Mahdi:16}.

In this paper, we study the influence of sterile neutrino dark matter particles 
on the process of \emph{reionization}, see e.g. extensive 
reviews~\cite{Barkana:00,Fan:06,Natarajan:14,Ferrara:14,McQuinn:15,Mesinger:16}. 
Generally, in models with non-zero initial velocities of dark matter particles 
the structure formation should be \emph{delayed} due to dark matter free streaming, see 
e.g. Fig.~3 of~\cite{Dayal:14}, causing a lack of HI ionizing photons, later end of reionization and
smaller CMB electron scattering optical depth $\tau_{\text{es}}$. On the other hand, because the 
haloes with the smallest mass could significantly contribute to recombination process effectively 
reducing the number of ionizing photons, 
see e.g.~\cite{Haiman:00,Barkana:02,Shapiro:03,Iliev:04a,Ciardi:05,McQuinn:06,Alvarez:10,Park:16}, 
one may expect reionization to be completed at \emph{earlier} times 
due to shortage of such smaller mass haloes in models with 
non-zero initial dark matter velocities. 
The effect has been first found in~\cite{Yue:12}, see 
also~\cite{Barkana:01,Somerville:03,Yoshida:03,Schultz:14,Dayal:15} for study of reionization in 
dark matter models in form of thermal relics with the mass of keV range. 
In their Fig.~2, \cite{Yue:12} showed that the 
reionization in thermal relic dark matter model with 10~keV mass finished earlier than in cold dark 
matter scenario, and that the process of reionization becomes sharper (in units of redshift) for warm 
dark matter compared to the CDM case. The results of these numerous studies, however, cannot be directly 
applied to study the process of reionization in realistic sterile neutrino dark matter models having 
non-thermal initial velocity distribution of dark matter particles~\cite{Shaposhnikov:08,Laine:08}. 
The goal of this paper is to extend the findings of~\cite{Yue:12} for the first time
to the case of sterile neutrino dark matter 
able to reproduce the observed emission line at $\sim$3.5~keV, and to compare them with
the up-to-date theoretical and observational uncertainties.

\section{Methods}\label{sec:methods}

To describe the process of reionization, we extended 
the `bubble model' formalism of~\cite{Furlanetto:04a} in a way close to~\cite{Yue:12}
who analyzed reionization for dark matter models in form of thermal relics with masses 2 and 10~keV. 
We assume that the main source of ionizing photons are Pop~II stars formed in 
galaxy-size haloes and consider mini-haloes only as additional `sinks' of ionizing radiation.

In the original bubble model formalism~\cite{Furlanetto:04a}, the mass of gas ionized by Pop~II stars 
$m_{\text{ion}}$, the mass of recombined hydrogen $m_{\text{rec}}$ and the mass of baryons 
collapsed into galaxies $m_{\text{gal}}$
are related with the simple \emph{linear} expression
\eq{\zeta m_{\text{gal}} = m_{\text{ion}} + m_{\text{rec}},\label{eq:bubble-int}}
where $\zeta$ is the number of ionizing photons per baryon released due to star formation during 
the process of halo collapse. 

We consider the following parametrization for $\zeta$ similar to~\cite{Yue:12}:
\eq{\zeta = f_* \times N_{\gamma/b} \times f_{\text{esc}},} 
where $f_*$, $N_{\gamma/b}$, $f_{\text{esc}}$
 are the star formation efficiency, the number of ionizing photons emitted per 
baryon in stars, and the fraction of ionizing photons escaping from galaxies.
Unlike e.g.~\cite{Inoue:06,Razoumov:06,Faisst:16}, 
we assume that these parameters remain constant with redshift.

Our fiducial value $\zeta = 15$ is chosen close to the most probable value of~\cite{Mitra:15}
which gives $\zeta \simeq 10-40$ based on $N_{\gamma/b} = 3200$. However, there is a large spread
in the literature among astrophysical parameters coming to $\zeta$.
For example, Ref.~\cite{Furlanetto:04a} used $f_{\text{esc}} = 0.2$, $f_* = 0.05$, $N_{\gamma/b} = 3200$,
which results in $\zeta = 32$.
According to Sec.~2.3 of~\cite{Iliev:04b}, for Pop~II low metallicity stars with a Salpeter 
IMF $N_{\gamma/b} = 3\ 000 - 10\ 000$~\cite{Leitherer:99}, which gives $\zeta = 30-100$ for 
$f_{\text{esc}}f_* = 0.01$.
Ref.~\cite{Becker:15} proposed the fiducial value $\zeta = 25$ showing illustrative values for its constituents: 
$N_{\gamma/b} = 4\ 648 (17\ 553)$ for a Salpeter (top-heavy) IMF~\cite{Wyithe:09},
$f_* = 0.01-0.1$~\cite{Behroozi:12} and $f_{\text{esc}} = 0.05-0.5$~\cite{Wise:14}.
Ref.~\cite{Siana:15} found $f_{\text{esc}} = 0.07-0.09$ for galaxies with $z > 3.06$.
Ref.~\cite{Grazian:15} constrained $f_{\text{esc}} < 0.02$ for their averaged sample 
of galaxies at $z\sim 3.3$.
Ref.~\cite{Izotov:16} observed a low-mass star-forming galaxy J0925$+$1403 at $z = 0.301$ 
having $f_{\text{esc}} = 0.078 \pm 0.011$.
Ref.~\cite{Borthakur:14} found $f_{\text{esc}} = 0.21 \pm 0.05$ for a starburst galaxy J0921$+$4509 
at $z = 0.23499$.
Ref.~\cite{Ma:16} found the significant increase of $f_{\text{esc}}$ (up to 20\% and above) by studying 
simulated feedback from stellar binary systems.
To reasonably account for all these uncertainties, we used $\zeta = 5$ and $\zeta = 45$ 
as our extreme values.

To collapse, a `bubble' region is assumed to contain mass in stars enough to ionize \emph{all} of its 
hydrogen atoms, so that the collapsed fraction $f_{\text{coll}}$ should not be smaller than the ratio 
of $m_{\text{gal}}/(m_{\text{ion}}+m_{\text{rec}})$: 
\eq{\zeta f_{\text{coll}} \geq 1 + \xi f_{\text{rec}}, \label{eq:bubble-main}}
where $f_{\text{rec}} = m_{\text{rec}}/(\xi\ m_{\text{ion}})$,
$\xi$ is the average number of recombinations per atom in collapsed mini-haloes
during the whole epoch of reionization.

We determine $f_{\text{coll}}$ and $f_{\text{rec}}$ as functions of halo mass $m$, 
redshift $z$ and mass overdensity $\delta_x$ by using the extended Press-Schechter 
formalism~\cite{Press:74,Bond:91,Bower:91,Lacey:93}:
\eq{
\begin{aligned}
f_{\text{coll}} = \mathcal{E}\left(m,\delta_x|m_{\text{min}},\delta_c/D(z)\right), \\
f_{\text{rec}} = \mathcal{E}\left(m,\delta_x|m_{\text{J}},\delta_c/D(z)\right)-
\mathcal{E}\left(m,\delta_x|m_{\text{min}},\delta_c/D(z)\right),
\label{eq:fcoll-eps}
\end{aligned}
}
where \eq{\mathcal{E}(m,\delta_x|m_i,\delta_c/D(z)) = \text{erfc}\left(\frac{\delta_c/D(z) - \delta_x}
 {\sqrt{2\left(\sigma^2(m_i)-\sigma^2(m)\right)}} \right),}
$\delta_c \approx 1.676$ for a flat $\Lambda$CDM model\footnote{Throughout 
this paper we used the latest best-fit cosmological parameters of \emph{Planck} 
satellite~\cite{Planck2015-XIII}: $\Omega_0 = 0.307$, $h = 0.678$, $\Omega_\Lambda = 0.693$, 
$\Omega_b = 0.0483$, $\sigma_8 = 0.823$ and $n_s = 0.961$.}~\cite{Eke:96},
$D(z)$ is the perturbation growth function~\cite{Heath:77},
\eq{D(z) = D_1(z)/D_1(0),\;\; D_1(z) = \frac{5\Omega_0}{2}g(z)\int\limits_z^\infty dz' \frac{1+z'}{g(z')^{3}}, \;\;
g^2(z) = \Omega_0(1+z)^3 + \Omega_\Lambda}
(see also~\cite{Eisenstein:97a} for an exact analytic expression for $D_1$), $\sigma^2(m)$ is 
 expressed through power spectrum $P(k)$\footnote{Here, $P(k)$ encodes the information
 about properties of dark matter particles, e.g. parameters $M_s$, $\sin^2(2\theta)$ and the value of lepton 
 asymmetry, see~\protect\cite{Laine:08,Abazajian:14,Venumadhav:15,Ghiglieri:15} 
 for detailed calculations used in our paper.} using
sharp $k$-space filter similar 
to~\cite{Bertschinger:06,Benson:12,Schneider:13,Schneider:14,Lovell:15,Schneider:16}:
\eq{\sigma^2(m) = \frac{1}{2\pi^2}\int\limits_0^{k_s}k^2 P(k) dk,\,
k_s = a \left(\frac{4\pi\bar\rho}{3m}\right)^{1/3},}
$a = 2.7$~\cite{Lovell:15}, $\bar\rho = \Omega_0 \rho_{crit}$ 
is the present-day average mass density in the Universe.

The minimal mass of collapsing haloes \eq{m_{\text{min}} = 3.59\times 10^7\ h^{-1}~\unit{M_\odot} 
\left(\frac{0.6}{\mu}\right)^{3/2}\left(\frac{10}{1+z}\right)^{3/2}}
 is taken from an assumption~\cite{Haiman:96} that the clouds will not fragment into new stars 
 until their virial temperature is higher than $T_{vir} > 10^4$~K, so the haloes can be 
 cooled by atomic hydrogen, according to e.g. Fig.~12 and Eq.~(26) of~\cite{Barkana:00}.
 Here we use mean molecular weight $\mu = 0.6$ as the average of $\mu = 0.59$ for a fully ionized primordial 
 gas and $\mu = 0.61$ for a gas with ionized hydrogen but only singly ionized helium, see Sec.~3.3 
 of~\cite{Loeb:10book}.

The minimal mass of mini-haloes responsible for recombinations (assumed to take place with the average number 
$\xi$ per atom) is the Jeans mass,
\eq{m_{\text{J}} = 3.96\times 10^3\ h^{-1}~\unit{M_\odot}
 \left(\frac{0.307}{\Omega_0}\right)^{1/2}
 \left(\frac{0.0222}{\Omega_b h^2}\right)^{3/5} 
\left(\frac{1+z}{10}\right)^{3/2}}
see Eq.~(41) of~\cite{Barkana:00}.
Our fiducial value of $\xi = 5$ and bounding values $\xi = 1$ and $\xi = 9$ were taken 
consistent with simulations of~\cite{Shapiro:03,Iliev:04a}\footnote{Note that our definition of $\xi$ is different from that of~\protect\cite{Iliev:04a}:
$\xi = \xi_{\text{Iliev}}-1$.}, see also Fig.~1 of~\cite{Finlator:12} as an illustration of apparent uncertainty.

By solving numerically Eq.~(\ref{eq:bubble-int}) using an approximation Eq.~(\ref{eq:fcoll-eps}) we obtained 
$\delta_x (z,\sigma^2(m))$. Consider then a linear barrier approximation
$\delta_x \approx B(m) \equiv B_0 + B_1\sigma^2(m)$ where $B_0$ and $B_1$ are calculated at the infinite masses 
(so that $\sigma^2 = 0$). In this case, an analytic expression for the mass function
is~\cite{Sheth:98}:\eq{m\frac{dn}{dm} = \sqrt{\frac{2}{\pi}}\frac{\bar\rho}{m}\left|\frac{d\ \log\sigma}{d\ \log m}
\right|
\frac{B_0}{\sigma(m)}\exp\left[-\frac{B^2(m)}{2\sigma^2(m)}\right].}
The volume filling fraction $Q_{\text{II}}(z)$ is then calculated:
\eq{Q_{\text{II}}(z) = \int \frac{m}{\bar\rho}\frac{dn}{dm}dm.}

 Motivated by recent study of~\cite{Furlanetto:15b} we define the length of reionization 
$\Delta z_{\text{rei}}$ as difference between the redshifts with volume fractions 0.9 and 0.1: 
\eq{\Delta z_{\text{rei}} = z\left(Q_{\text{II}} = 0.1\right) - z\left(Q_{\text{II}} = 0.9\right).
\label{eq:deltazrei}}
We also assume that the reionization has been completed at $z_{\text{rei}} = z\left(Q_{\text{II}} = 0.99\right)$.

To calculate the CMB electron scattering optical depth $\tau_{\text{es}}(z)$ at redshift $z$, we used Eq.~(9) 
of~\cite{Dayal:15}
\begin{equation}
\label{eq:tau}
\tau_{\text{es}}(z) = c \bar n_{\text H} \sigma_{\text T} \int_{0}^{z} f_e(z') Q_{\text{II}}(z') \frac{(1+z')^{2}}{H(z')} dz',
\end{equation}
where $c$ is the speed of light, $\bar n_{\text H}$ is the mean comoving number density of hydrogen atoms, 
$\sigma_{\text T}$ is the Thomson cross section, $f_e(z)$ is the number of free electrons per hydrogen atom
calculated under assumption of~\cite{Kuhlen:12}, 
and $H(z)$ is the Hubble parameter. 

\section{Results and conclusions}

We analyze the process of reionization in several dark matter models, including cold dark matter
and several sterile neutrino dark matter models depicted in
Fig.~4 of~\cite{Lovell:15} calculated using approach of~\cite{Laine:08,Ghiglieri:15},
as well as sterile neutrino dark models from 
Fig.~2 of~\cite{Horiuchi:15} calculated using another approach~\cite{Abazajian:14,Venumadhav:15}.

The models of~\cite{Lovell:15} are parametrized by the value of lepton asymmetry 
\eq{L_6 \equiv 10^6\ \frac{n_{\nu_e}-n_{\bar\nu_e}}{s},\label{eq:l6-definition}}
where $n_{\nu_e}$ and $n_{\bar\nu_e}$ are the number densities of electron neutrinos and antineutrinos, 
$s$ is the total entropy density.
For a sterile neutrino dark matter with 
the mass of 7~keV, the value of $L_6$ can be translated to sterile neutrino mixing angle (Eq.~\ref{eq:m-theta}),
according to Fig.~1 of~\cite{Lovell:15}. 
For example, the region for $\sin^2(2\theta) \simeq (4.5-6)\times 10^{-11}$ 
preferred by the $\sim$3.5~keV line measurements~\cite{Boyarsky:14b,Iakubovskyi:15b,Ruchayskiy:15}
corresponds to $L_6 \simeq 9-10$. 
As a result, we consider the model with $L_6 = 10$ as our fiducial model,
in addition to the model s228899 of~\cite{Horiuchi:15} which corresponds 
to $\sin^2(2\theta) = 2.8899 \times 10^{-11}$.
The values of $L_6 \gtrsim 10$ would produce smaller mixing angles and therefore cannot explain the 
observed $\sim$3.5~keV line but are not excluded by existing measurements, 
so we consider them for illustrative purposes, as well as the model s208 of~\cite{Horiuchi:15} 
which corresponds to $\sin^2(2\theta) = 0.8 \times 10^{-11}$.

We summarize the obtained results in Table~\ref{tab:results} where
we mark in bold our fiducial models that match the \emph{existing} observational constraints 
($z_{\text{rei}} \gtrsim 5.6$, $\tau_{\text{es}} = 0.046-0.103$), see
also detailed discussion in~\cite{Robertson:13,Robertson:15}:
\begin{itemize}
 \item Lyman-$\alpha$ forest transmission~\cite{Fan:05} requires $Q_{\text{II}} \simeq 0.96-0.99$ at $z = 6.2$;
 \item dark Lyman-$\alpha$ pixels~\cite{McGreer:14} suggests $1\sigma$ upper bounds: 
 $Q_{\text{II}} > 0.89$ at $z = 5.9$, $Q_{\text{II}} > 0.91$ at $z = 5.5$;
 \item the presence of Gunn-Peterson damping wings in high-$z$ quasar spectra, see 
 e.g.~\cite{Bolton:11,Schroeder:12} (see however~\cite{Bosman:15}) suggests \emph{upper} bound
 $Q_{\text{II}} \lesssim 0.97$ at $z \approx 6$;
 \item the absence of Gunn-Peterson damping wings in high-$z$ GRB spectra
 implies $2\sigma$ bound $Q_{\text{II}} \geq 0.89$ at $z = 5.913$~\cite{Chornock:13};
 \item the redshift decline of the Lyman-$\alpha$ emitter luminosity function  
 suggests $Q_{\text{II}} \sim 0.58-0.88$ at $z = 6.6$ and 
 $Q_{\text{II}} \sim 0.46-0.88$ at $z = 7.0$~\cite{Ota:07}, and $Q_{\text{II}} \gtrsim 0.5-0.6$ 
 (with the strongest model-dependent constraint $Q_{\text{II}} \gtrsim 0.8 \pm 0.2$) 
 at $z = 6.6$~\cite{Ouchi:10};
 \item electron scattering optical depth $\tau_{\text{es}}$ inferred from CMB measurements. 
 Throughout this paper, 
 we used the latest values $\tau_{\text{es}} = 0.058 \pm 0.012$ recently reported by 
 \emph{Planck} collaboration~\cite{Planck2016-XLVII} for their combined analysis 
 of \emph{Planck} E-mode polarization correlations at low-$l$ (\texttt{lollipop}) 
 and \emph{Planck} temperature-temperature (TT) correlations.
However, we should also be aware of the recent estimate $\tau_{\text{es}} = 0.055 \pm 0.009$ 
(with $2\sigma$ upper bound of $0.073$) based on \texttt{SimBal} 
likelihood estimate from cross-correlating E-mode polarization maps at 100 and 143 
GHz~\cite{Planck2016-XLVI}, the previous release of \emph{Planck} data resulting in
 $\tau_{\text{es}} = 0.066 \pm 0.016$~\cite{Planck2015-XIII}, and of previous experiment \emph{WMAP}
which gives $\tau_{\text{es}} = 0.089 \pm 0.014$~\cite{Hinshaw:12}.
 \end{itemize}

\begin{table*}
\centering
\begin{tabular}[c]{lcccccl}
\hline
Model & $\zeta$ & $\xi$ & $z_{\text{rei}}$ & $\Delta z_{\text{rei}}$ & $\tau_{\text{es}}$ & Comments \\
\hline
\multicolumn{7}{c}{Cold dark matter (CDM) model:} \\
\hline
\textbf{CDM-15-5} & \textbf{15} & \textbf{5} & \textbf{6.7} & \textbf{7.4} & \textbf{0.084} & \textbf{Fiducial model for \emph{Planck15}} \\ 
CDM-11-5 & 11 & 5 & 5.6 & 8.1 & 0.073 & Model with smaller $\zeta$ to reconcile with~\protect\cite{Planck2016-XLVII} \\ 
CDM-5-5 & 5 & 5 & 2.8 & 11.3 & 0.063 & Lower $\zeta$, excluded by too small $z_{\text{rei}}$ \\
CDM-45-5 & 45 & 5 & 10.4 & 5.9 & 0.119 & Upper $\zeta$, in tension with~\cite{Bolton:11,Schroeder:12,Planck2015-XIII,Planck2016-XLVII} \\ 
CDM-15-0 & 15 & 0 & 9.0 & 7.5 & 0.109 & Fiducial model for \emph{Planck15} with $\xi = 0$ \\
CDM-15-9 & 15 & 9 & 5.4 & 7.8 & 0.072 & Upper $\xi$, excluded by too small $z_{\text{rei}}$ \\ 
CDM-15-1 & 15 & 1 & 8.5 & 7.5 & 0.103 & Lower $\xi$, in tension with~\cite{Bolton:11,Schroeder:12,Planck2015-XIII,Planck2016-XLVII} \\ 
CDM-45-9 & 45 & 9 & 9.4 & 5.9 & 0.107 & Upper $\zeta$ and $\xi$, in tension with~\cite{Bolton:11,Schroeder:12,Planck2015-XIII,Planck2016-XLVII} \\ 
CDM-5-1 & 5 & 1 & 4.9 & 10.5 & 0.078 & Lower $\zeta$ and $\xi$, excluded by too small $z_{\text{rei}}$ \\ 
\hline
\multicolumn{7}{c}{Sterile neutrino dark matter models from~\protect\cite{Lovell:15}:} \\
\hline
L8-15-5 & 15 & 5 & 7.5 & 5.8 & 0.084 &  \\ 
\textbf{L10-15-5} & \textbf{15} & \textbf{5} & \textbf{6.9} & \textbf{5.2} & \textbf{0.074} & \textbf{Fiducial model for \emph{Planck15}} \\ 
L10-11-5 & 11 & 5 & 6.3 & 5.6 & 0.070 & Model with smaller $\zeta$ to reconcile with~\protect\cite{Planck2016-XLVII} \\ 
L16-15-5 & 15 & 5 & 6.2 & 4.7 & 0.064 &  \\ 
L20-15-5 & 15 & 5 & 6.1 & 4.7 & 0.063 &  \\ 
L50-15-5 & 15 & 5 & 5.9 & 4.6 & 0.060 &  \\ 
L120-15-5 & 15 & 5 & 5.8 & 4.5 & 0.060 &  \\ 
L700-15-5 & 15 & 5 & 5.8 & 4.5 & 0.059 &  \\ 
\hline
\multicolumn{7}{c}{Sterile neutrino dark matter models from~\protect\cite{Horiuchi:15}:} \\
\hline
s220-15-5 & 15 & 5 & 7.0 & 5.4 & 0.076 &  \\ 
\textbf{s228899-15-5} & \textbf{15} & \textbf{5} & \textbf{6.3} & \textbf{4.8} & \textbf{0.066} & \textbf{Fiducial model for \emph{Planck15}} \\
s228899-11-5 & 11 & 5 & 5.8 & 5.2 & 0.062 & Model with smaller $\zeta$ to reconcile with~\protect\cite{Planck2016-XLVII} \\ 
s208-15-5 & 15 & 5 & 5.8 & 4.4 & 0.059 & \\ 
\hline
\end{tabular}
\caption{The values of $z_{\text{rei}} = z\left(Q_{\text{II} = 0.99}\right)$, 
$\Delta z_{\text{rei}}$ (Eq.~(\protect\ref{eq:deltazrei}))
and $\tau_{\text{es}}$ (Eq.~(\protect\ref{eq:tau})) for reionization models considered in this paper.
For models starting with L, the first number denotes the value of electron lepton asymmetry $L_6$ 
according to Eq.~\protect\ref{eq:l6-definition};
for models staring with s2, the first number denotes
sterile neutrino mixing angle $\sin^2(2\theta)$ in units of $10^{-11}$ according to Eq.~\protect\ref{eq:m-theta}.
Our model parameters $\zeta$ and $\xi$ denote dimensionless ionization efficiency 
(the number of UV ionizing photons produced 
per baryon in haloes released in the process of halo collapse) and 
minihalo recombination efficiency (the number of recombinations per atom in collapsed mini-haloes), see 
Sec.~\protect\ref{sec:methods}. The models CDM-5-5 and CDM-5-1 are excluded having reionization
finished too late; on the other hand, the models CDM-45-5, CDM-15-1 and CDM-45-9 are in tension with 
\emph{Planck} measurements of $\tau_{\text{es}}$~\protect\cite{Planck2015-XIII,Planck2016-XLVII} 
and have reionization too early contrary to
\emph{upper} bounds on $Q_{\text{II}}$ obtained by~\protect\cite{Bolton:11,Schroeder:12}.
In addition to CDM models, 
we show the behavior of sterile neutrino dark matter models L10-15-5 of~\protect\cite{Lovell:15} and 
s228899-15-5 of~\protect\cite{Horiuchi:15} as our fiducial models able to reproduce the observed emission line
at $\sim$3.5~keV, see~\protect\cite{Boyarsky:14b,Iakubovskyi:15b,Ruchayskiy:15} for details.
Other sterile neutrino dark matter models are presented for illustrative purposes.
The obtained change in $z_{\text{rei}}$ (up to 0.4) and $\tau_{\text{es}}$ (up to 0.018) 
between these models and the cold dark matter model CDM-15-5 is below the existing uncertainties on 
reionization efficiency $\zeta$ and the possible QSO contribution~\protect\cite{Madau:15} 
but is comparable with e.g. possible contribution of Pop~III 
stars~\protect\cite{Dixon:15} and Lyman-limit systems~\protect\cite{Shukla:16}. 
More important is the systematic decrease in $\Delta z_{\text{rei}}$ (up to 2.6).
In all models, matter power spectra were calculated by using the latest best-fit cosmological parameters
obtained by \emph{Planck} satellite, see~\protect\cite{Planck2015-XIII}.}
\label{tab:results}
\end{table*}

\begin{figure*}[tp!]
   \includegraphics[width=0.49\textwidth]{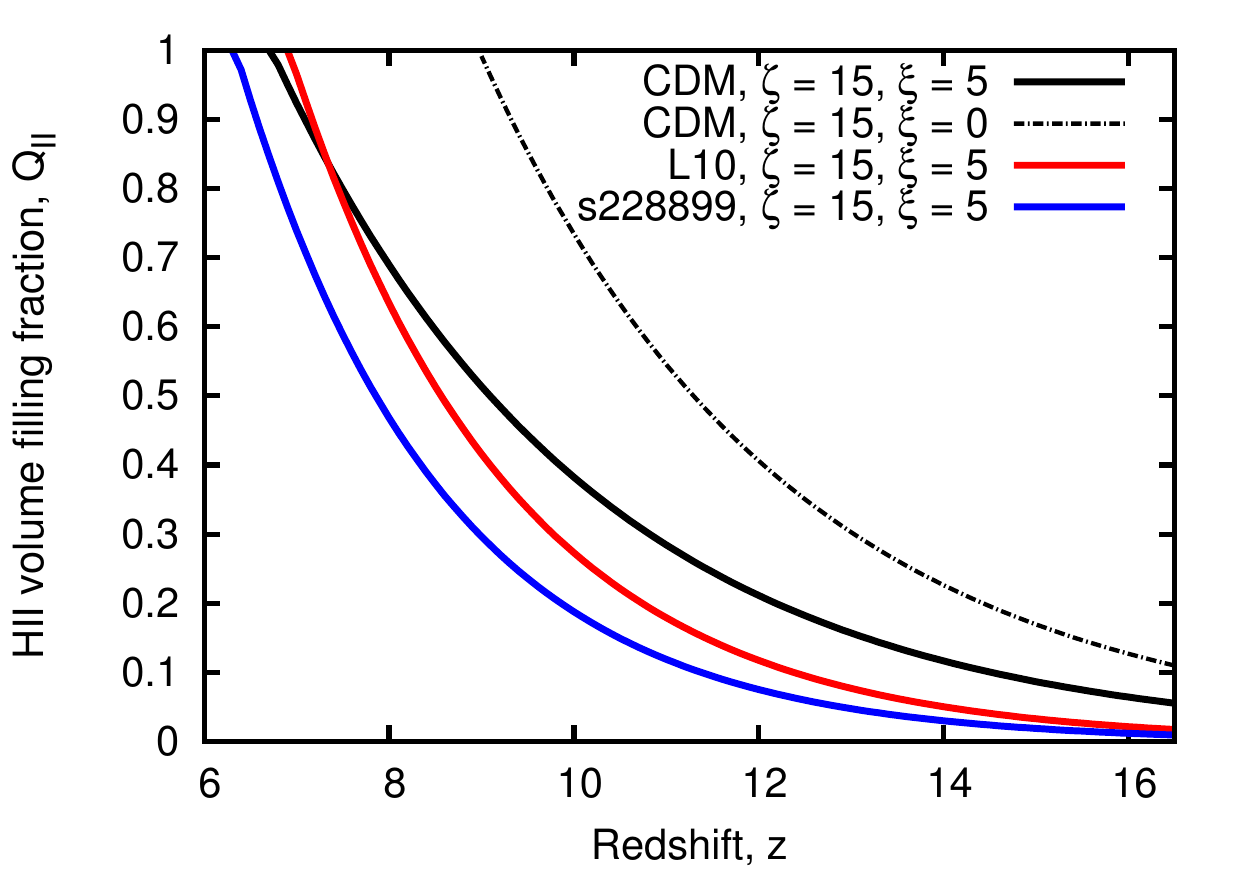}~%
   \includegraphics[width=0.49\textwidth]{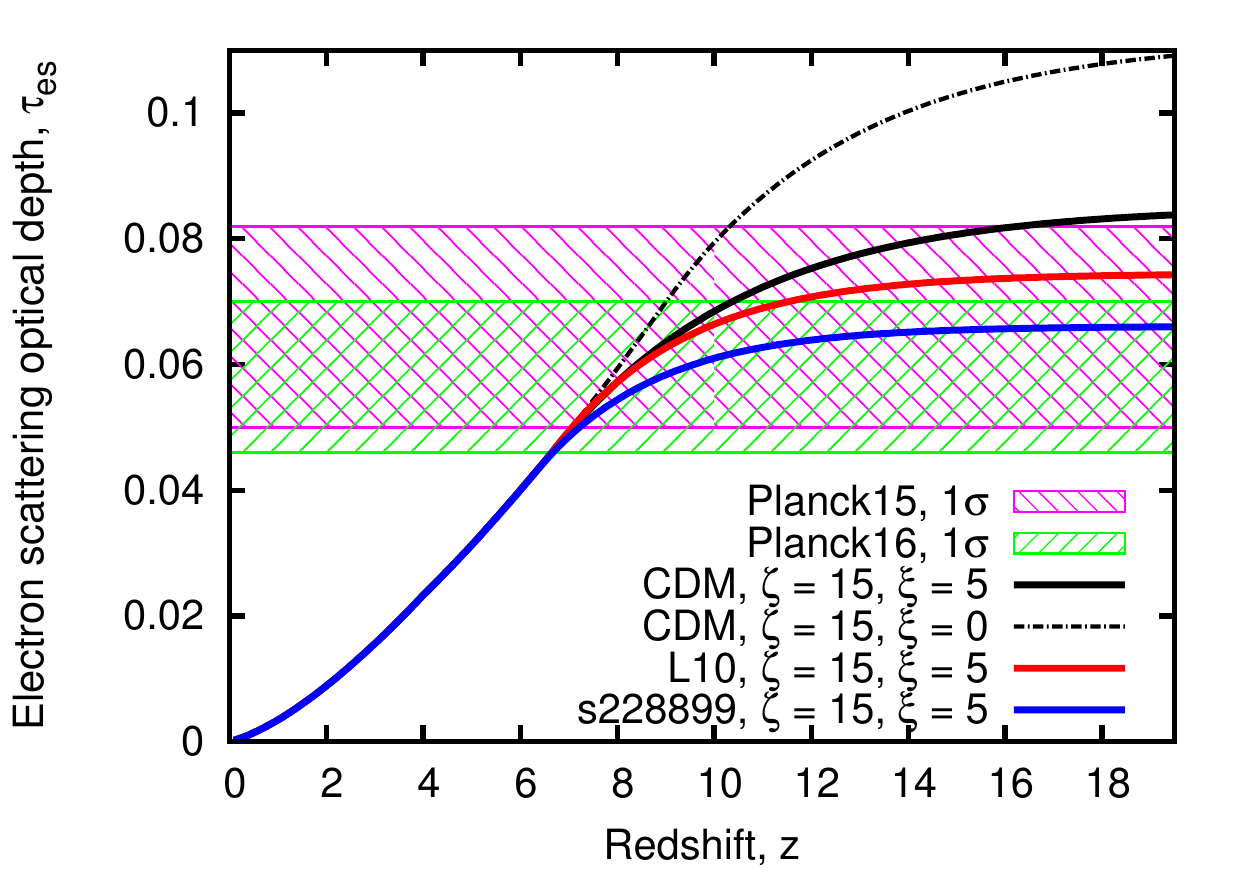}
  \caption{\textit{Left}: dependence of the HII volume filling fraction $Q_{\text{II}}$ on redshift $z$ 
  for our fiducial dark matter models, see Table~\protect\ref{tab:results}. 
  L10 indicates the value of electron lepton asymmetry $L_6 = 10$ 
  for models of~\protect\cite{Lovell:15} while s228899 
  indicates the values of $\sin^2(2\theta) = 2.8899\times 10^{-11}$
  for models of~\protect\cite{Horiuchi:15}. 
  Here one may see that in L10 model the reionization is to be completed at earlier times 
  compared to CDM scenario; 
  despite the decrease of halo number density in L10 model should \emph{delay}
  the reionization, this is overcompensated by the sufficient decrease in recombinations, see the 
  back dashed line for the same CDM model \emph{without} recombinations. 
  \textit{Right}: The same for dependence of electron scattering optical depth $\tau_{\text{es}}$ on 
  redshift $z$, plotted against the latest measurements by \emph{Planck} 
  satellite~\protect\cite{Planck2015-XIII,Planck2016-XLVII}. 
  Although the models CDM-15-5 and L10-15-5 depicted here become 
  offset from 1$\sigma$ range of $\tau_{\text{es}}$ obtained by~\protect\cite{Planck2016-XLVII}, 
  Table~\protect\ref{tab:results} shows that this difference can be adjusted with a
  slight change of ionization efficiency $\zeta$, much smaller than allowed by existing 
  observations.
  }
  \label{fig:z-tau}
\end{figure*}

Fig.~\ref{fig:z-tau} illustrates
evolution histories for $Q_{\text{II}}$ and $\tau_{\text{es}}$ for our fiducial models.
Interestingly, in our sterile neutrino dark matter models L8-15-5 and L10-15-5 
reionization completed at earlier times compared to the corresponding cold dark matter model CDM-15-5; 
despite the decrease of halo number density in sterile neutrino dark matter models 
should delay the reionization compared to CDM scenario, 
this is overcompensated by the corresponding shortage of recombinations (see the model CDM-15-0 for details). 
We should stress, however, that the fact that in the models L8-15-5 and L10-15-5 reionization ends close 
to that in the model CDM-15-5 is a pure numerical coincidence between smaller halo suppression 
and the fiducial values of model parameters $\zeta$ and $\xi$. For example, smaller values of $\xi$
adopted in a subsequent paper~\cite{Bose:16b} leads to qualitatively different behaviour of $z_{\text rei}$: 
namely, in sterile neutrino dark matter models reionization ends later than in CDM.
We discuss this in more details at the end of Sec.~\ref{sec:discussion}.

The existing uncertainty in halo ionization efficiency $\zeta$ 
 has the largest impact on reionization parameters $z_{\text{rei}}$ (variations up to 7.6) 
 and $\tau_{\text{es}}$ (variations up to 0.056), although these variations are well-correlated. 
 The impact of this uncertainty is much above the expected influence of Pop~III stars~\cite{Dixon:15}
 and of Lyman-limit systems~\cite{Shukla:16}, and is comparable with the maximal estimated influence 
 of high-$z$ quasars~\cite{Madau:15}.
 The influence on the length of reionization $\Delta z_{\text{rei}}$ is smaller (variations up to 5.4) 
 and is also strongly correlated with the changes of $z_{\text{rei}}$ and $\tau_{\text{es}}$.

 The influence of the uncertainty in minihalo recombination efficiency $\xi$ is 
the largest in CDM models inducing minor difference of 
 $z_{\text{rei}}$ (variations up to 3.1) and $\tau_{\text{es}}$ (variations up to 0.031) 
 with much smaller influence on $\Delta z_{\text{rei}}$ (variations up to 0.4). 
 For sterile neutrino dark matter models,
 the influence is by a factor of $\gtrsim 20$ smaller due to strong suppression of mini-haloes 
 and thus can be neglected.
 
 Initial velocities of dark matter particles imposed by 
 sterile neutrino dark matter models with parameters consistent with observation of the new line at 
 $\sim$3.5~keV (our fiducial models in Table~\ref{tab:results}) 
 introduce a modest influence on $z_{\text{rei}}$ (variations up to 0.4) and $\tau_{\text{es}}$ 
 (variations up to 0.018), close to the expected influence of Pop~III stars~\cite{Dixon:15}
 and of Lyman-limit systems~\cite{Shukla:16}.
 Due to much larger influence of $\zeta$ on $z_{\text{rei}}$ and $\tau_{\text{es}}$, it is not possible to
 prefer or rule out the selected dark matter model based on existing observational constrains on these 
 parameters. The influence of sterile neutrino dark matter on $\Delta z_{\text{rei}}$, however, 
 is much more profound (variations by 1.6-2.6). Moreover, it is impossible to `imitate'
 the effect of sterile neutrino dark matter on the entire set of parameters 
 $\{z_{\text{rei}}, \Delta z_{\text{rei}}, \tau_{\text{es}}\}$ with \emph{any} 
 reasonable value of $\zeta$ and $\xi$ for CDM models, as Table~\ref{tab:results} demonstrates.

\section{Discussion}\label{sec:discussion}

Throughout this work we assume that the dominant sources of ionizing photons are Pop~II stars 
forming at galaxies. Although such an assumption is often considered to be conventional, 
see e.g.~\cite{Barkana:00}, it by no means excludes possible significant contribution of other ionizing sources.
For example, we do not take into account the ionizing UV emission from Pop~III stars, 
because their contribution is very uncertain and is thought to be 
subdominant~\cite{Wise:12,Johnson:12,Paardekooper:12}, 
see however~\cite{Yajima:14,Wise:14}.
Recent detailed simulations of~\cite{Dixon:15} show that UV radiation from Pop~III stars advance 
reionization by $\Delta z \leq 1.3$ and increase $\tau_{\text{es}}$ 
by $\Delta \tau_{\text{es}} \leq 0.019$, comparable with other uncertainties.
Similarly, we do not take into account Lyman-limit systems; 
Ref.~\cite{Shukla:16} found that they delay reionization
by $\Delta z = 0.6-0.8$ with small (by $\sim 0.002$) decrease of $\tau_{\text{es}}$.
Also, we do not take into account UV radiation from active galactic nuclei (AGN) because of their suppressed 
contribution at high redshifts shown in e.g.~\cite{Mitra:15}; although~\cite{Madau:15} showed that it is even
possible to reproduce reionization using AGN as only sources for reionization using new measurement of 
high-$z$ AGN abundance~\cite{Giallongo:15}, 
but this hypothesis is in apparent tension with recent observations of metal absorbers at 
$z\sim 6$~\cite{Finlator:16}.

The presence of X-ray ionizing radiation may further complicate the description of reionization, because X-rays 
can easily escape UV bubbles due to their much larger mean free path compared to UV photons. 
Notably, they may originate in quantities required for reionization 
not only from conventional astrophysical sources such as X-ray binaries and accreting massive black holes,
see e.g.~\cite{McQuinn:12}, but also from decays of dark matter particles such as sterile neutrinos,
see e.g.~\cite{Dolgov:00,Abazajian:01a,Hansen:03,Pierpaoli:03,Liu:16,Oldengott:16}.
In addition, dark matter decays could broaden the CMB last scattering surface and could have an imprint on CMB 
anisotropies~\cite{Adams:98,Chen:03,Padmanabhan:05}. 
Another effect of X-rays (e.g. from decaying dark matter) 
catalyzing the formation of molecular hydrogen that may speed up gas cooling and star 
formation suggested by~\cite{Biermann:06} is later found negligible~\cite{Ripamonti:06b}.
It is important to note, however, that the obtained upper limits on dark matter radiative decay lifetime
are about two orders of magnitude weaker than the expected decay lifetime of sterile neutrino dark matter
responsible for $\sim$3.5~keV line, see e.g. Fig.~10 of~\cite{Oldengott:16}.
This validates neglecting the effects of dark matter decays throughout this paper.

To further \emph{probe} the influence of sterile neutrino dark matter on reionization,
one needs more detailed observations sensitive to $\Delta z_{\text{rei}}$, in addition
to existing probes of $z_{\text{rei}}$ and $\tau_{\text{es}}$.
Future measurements of 21-cm line fluctuations (see e.g.~\cite{Sitwell:13,Furlanetto:15a,Liu:15,Fialkov:16} 
and references therein) and 
more detailed study of kinetic Sunyaev-Zeldovich (kSZ) effect (see e.g. 
Ref.~\cite{Park:13,Calabrese:14,Alvarez:15} 
and references therein) are the straightforward candidates for such an analysis. For example, the
value of $\Delta z_{\text{rei}}$ obtained in our preferred sterile neutrino dark matter models
tends to be in better consistency with recent kSZ constraints~\cite{George:14,Planck2016-XLVII}, 
compared with the standard $\Lambda$CDM scenario.

Our work is based on the extension of the initial formalism of~\cite{Furlanetto:04a}.
Recently, several physically motivated semi-analytical models have been proposed, 
based on excursion set peaks formalism~\cite{Paranjape:14},
percolation theory~\cite{Xu:13,Furlanetto:15b},
the specially adapted GALFORM~\cite{Cole:00,Lacey:15} 
model of galaxy formation~\cite{Gonzalez-Perez:13,Kennedy:13,Lovell:15},
and adjusting better consistency with up-to-date numerical simulations~\cite{Kaurov:15b}.
Because these extensions remain physically well-justified and computationally inexpensive
compared to detailed simulations 
(see e.g.~\cite{Schultz:14,Bose:15,Sarkar:15} for recent simulations with non-trivial dark matter power spectra),
their usage is well-motivated in a future extension of this work.

For example, it is informative to compare our results with the recent findings
of~\cite{Bose:16b} who used the GALFORM semi-analytic model of galaxy formation to model reionization
for CDM and several sterile neutrino dark matter models (including L8 and L700 models also analyzed 
in our Table~\ref{tab:results}). First of all, we must emphasize that the authors of~\cite{Bose:16b} 
have confirmed our major \emph{qualitative} result -- much sharper reionization in sterile neutrino 
dark matter models compared to $\Lambda$CDM. We also confirm the finding of~\cite{Bose:16b}
that the median halo mass responsible for reionization is much larger for sterile neutrino dark matter
models than that for $\Lambda$CDM (expected due to decrease of substantial 
smaller-mass haloes in sterile neutrino dark matter models). For example, the median halo mass where
the recombination is completed at $z = 10$ for 
the model CDM-15-5 is $0.8\times 10^9\ h^{-1}\ M_\odot$, while for the
model L10-15-5 it is $5.5\times 10^9\ h^{-1}\ M_\odot$ and for the model s228899-15-5 it is 
$1.2\times 10^{10}\ h^{-1}\ M_\odot$, in agreement with Fig.~5 of~\cite{Bose:16b}.
The fact that in our models L8-15-5 and L10-15-5 reionization 
finishes faster than in the CDM-15-5 model, contrary to findings of~\cite{Bose:16b}, 
see e.g. their footnote 2, is due to different description of recombination process. 
For example, the authors of~\cite{Bose:16b} assume the number of recombinations per baryon to be the same for
all halo masses, while in our approach it is the same only for minihaloes -- where the effect of recombinations
should be the most pronounced because of the largest baryon density -- which experience the 
largest dearth due to non-zero initial velocities of sterile neutrino dark matter. Assuming instead that 
the number of recombinations per baryon is the same for all haloes (which is, in terms of our model, 
a simple redefinition of $\zeta$ with taking $\xi$ to 0), we are able to reproduce 
the result of~\cite{Bose:16b} that in sterile neutrino dark matter models the reionization 
ends \emph{later} than in $\Lambda$CDM.
However, there is another notable difference between our findings and the results 
of~\cite{Bose:16b}. Namely, in their models (involving either $\Lambda$CDM or sterile neutrino dark matter)
the reionization evolves systematically faster 
(with $\Delta z_{\text{rei}}$ smaller by a factor of $\sim 2-3$) compared 
to our results, as well as the results of several other 
groups~\cite{Furlanetto:04a,Yue:12,McQuinn:12,Mitra:15,Shukla:16,Greig:16} 
(see however~\cite{Finlator:14,Ocvirk:15}). 
Successful resolution of this difference in $\Delta z_{\text{rei}}$ (which appears to be sensitive to the 
very details of reionization astrophysics) 
will be essential to quantify how much the recently obtained kSZ 
constraints~\cite{George:14,Planck2016-XLVII} are 
better consistent with sterile neutrino dark matter than with conventional $\Lambda$CDM scenario.

\acknowledgments

We thank K.~Abazajian and M.~Lovell for comments and for 
sharing power spectra for a set of sterile neutrino dark matter models used in this work. 
We also thank S.~Bose, S.~Cole, A.~Ferrara, S.~H.~Hansen, P.~Naselsky, O.~Ruchayskiy, S.~Sarkar
and the anonymous Referees for comments. 
The work of A.~R.\ was partially supported from the Swiss National Science Foundation 
grant SCOPE IZ7370-152581, the State Fund for Fundamental Research of Ukraine grant
F64/45-2016, and the Program of Cosmic Research of the National Academy of Sciences of Ukraine.
The work of D.~I. was supported by a research grant from VILLUM FONDEN.


\bibliographystyle{JHEP}%

\bibliography{preamble,reionization}

\let\jnlstyle=\rm\def\jref#1{{\jnlstyle#1}}\def\aj{\jref{AJ}}
  \def\araa{\jref{ARA\&A}} \def\apj{\jref{ApJ}} \def\apjl{\jref{ApJ}}
  \def\apjs{\jref{ApJS}} \def\ao{\jref{Appl.~Opt.}} \def\apss{\jref{Ap\&SS}}
  \def\aap{\jref{A\&A}} \def\aapr{\jref{A\&A~Rev.}} \def\aaps{\jref{A\&AS}}
  \def\azh{\jref{AZh}} \def\baas{\jref{BAAS}} \def\jcap{\jref{JCAP}}
  \def\jrasc{\jref{JRASC}} \def\memras{\jref{MmRAS}} \def\mnras{\jref{MNRAS}}
  \def\pra{\jref{Phys.~Rev.~A}} \def\prb{\jref{Phys.~Rev.~B}}
  \def\prc{\jref{Phys.~Rev.~C}} \def\prd{\jref{Phys.~Rev.~D}}
  \def\pre{\jref{Phys.~Rev.~E}} \def\prl{\jref{Phys.~Rev.~Lett.}}
  \def\pasa{\jref{PASA}} \def\pasp{\jref{PASP}} \def\pasj{\jref{PASJ}}
  \def\qjras{\jref{QJRAS}} \def\skytel{\jref{S\&T}}
  \def\solphys{\jref{Sol.~Phys.}} \def\sovast{\jref{Soviet~Ast.}}
  \def\ssr{\jref{Space~Sci.~Rev.}} \def\zap{\jref{ZAp}} \def\nat{\jref{Nature}}
  \def\iaucirc{\jref{IAU~Circ.}} \def\aplett{\jref{Astrophys.~Lett.}}
  \def\apspr{\jref{Astrophys.~Space~Phys.~Res.}}
  \def\bain{\jref{Bull.~Astron.~Inst.~Netherlands}}
  \def\fcp{\jref{Fund.~Cosmic~Phys.}} \def\gca{\jref{Geochim.~Cosmochim.~Acta}}
  \def\grl{\jref{Geophys.~Res.~Lett.}} \def\jcp{\jref{J.~Chem.~Phys.}}
  \def\jgr{\jref{J.~Geophys.~Res.}}
  \def\jqsrt{\jref{J.~Quant.~Spec.~Radiat.~Transf.}}
  \def\memsai{\jref{Mem.~Soc.~Astron.~Italiana}}
  \def\nphysa{\jref{Nucl.~Phys.~A}} \def\physrep{\jref{Phys.~Rep.}}
  \def\physscr{\jref{Phys.~Scr}} \def\planss{\jref{Planet.~Space~Sci.}}
  \def\procspie{\jref{Proc.~SPIE}} \let\astap=\aap \let\apjlett=\apjl
  \let\apjsupp=\apjs \let\applopt=\ao
\providecommand{\href}[2]{#2}\begingroup\raggedright\begin{thebibliography}{100}

\bibitem{Szalay:76}
A.~S. {Szalay} and G.~{Marx}, \emph{{Neutrino rest mass from cosmology}},
  {\emph{\aap} {\bf 49} (June, 1976) 437--441}.

\bibitem{Lee:77}
B.~W. {Lee} and S.~{Weinberg}, \emph{{Cosmological lower bound on
  heavy-neutrino masses}},
  \href{http://dx.doi.org/10.1103/PhysRevLett.39.165}{\emph{Physical Review
  Letters} {\bf 39} (July, 1977) 165--168}.

\bibitem{Goldberg:83}
H.~{Goldberg}, \emph{{Constraint on the photino mass from cosmology}},
  \href{http://dx.doi.org/10.1103/PhysRevLett.50.1419}{\emph{Physical Review
  Letters} {\bf 50} (May, 1983) 1419--1422}.

\bibitem{Ellis:84}
J.~{Ellis}, J.~S. {Hagelin}, D.~V. {Nanopoulos}, K.~{Olive} and M.~{Srednicki},
  \emph{{Supersymmetric relics from the big bang}},
  \href{http://dx.doi.org/10.1016/0550-3213(84)90461-9}{\emph{Nuclear Physics
  B} {\bf 238} (June, 1984) 453--476}.

\bibitem{Servant:02}
G.~{Servant} and T.~M.~P. {Tait}, \emph{{Is the lightest Kaluza-Klein particle
  a viable dark matter candidate?}},
  \href{http://dx.doi.org/10.1016/S0550-3213(02)01012-X}{\emph{Nuclear Physics
  B} {\bf 650} (Feb., 2003) 391--419},
  [\href{http://arxiv.org/abs/hep-ph/0206071}{{\tt hep-ph/0206071}}].

\bibitem{Marsh:15}
D.~J.~E. {Marsh}, \emph{{Axion Cosmology}}, {\emph{ArXiv e-prints} (Oct., 2015)
  }, [\href{http://arxiv.org/abs/1510.07633}{{\tt 1510.07633}}].

\bibitem{Preskill:83}
J.~{Preskill}, M.~B. {Wise} and F.~{Wilczek}, \emph{{Cosmology of the invisible
  axion}}, \href{http://dx.doi.org/10.1016/0370-2693(83)90637-8}{\emph{Physics
  Letters B} {\bf 120} (Jan., 1983) 127--132}.

\bibitem{Abbott:83}
L.~F. {Abbott} and P.~{Sikivie}, \emph{{A cosmological bound on the invisible
  axion}}, \href{http://dx.doi.org/10.1016/0370-2693(83)90638-X}{\emph{Physics
  Letters B} {\bf 120} (Jan., 1983) 133--136}.

\bibitem{Dine:83}
M.~{Dine} and W.~{Fischler}, \emph{{The not-so-harmless axion}},
  \href{http://dx.doi.org/10.1016/0370-2693(83)90639-1}{\emph{Physics Letters
  B} {\bf 120} (Jan., 1983) 137--141}.

\bibitem{Bertone:04}
G.~{Bertone}, D.~{Hooper} and J.~{Silk}, \emph{{Particle dark matter: evidence,
  candidates and constraints}},
  \href{http://dx.doi.org/10.1016/j.physrep.2004.08.031}{\emph{\physrep} {\bf
  405} (Jan., 2005) 279--390}, [\href{http://arxiv.org/abs/hep-ph/0404175}{{\tt
  hep-ph/0404175}}].

\bibitem{Bergstrom:09}
L.~{Bergstr{\"o}m}, \emph{{Dark matter candidates}},
  \href{http://dx.doi.org/10.1088/1367-2630/11/10/105006}{\emph{New Journal of
  Physics} {\bf 11} (Oct., 2009) 105006--+},
  [\href{http://arxiv.org/abs/0903.4849}{{\tt 0903.4849}}].

\bibitem{Feng:10}
J.~L. {Feng}, \emph{{Dark Matter Candidates from Particle Physics and Methods
  of Detection}},
  \href{http://dx.doi.org/10.1146/annurev-astro-082708-101659}{\emph{\araa}
  {\bf 48} (Sept., 2010) 495--545}, [\href{http://arxiv.org/abs/1003.0904}{{\tt
  1003.0904}}].

\bibitem{Drees:12}
M.~{Drees} and G.~{Gerbier}, \emph{{Mini--Review of Dark Matter: 2012}},
  {\emph{ArXiv e-prints} (Apr., 2012) },
  [\href{http://arxiv.org/abs/1204.2373}{{\tt 1204.2373}}].

\bibitem{Bertone:16}
G.~{Bertone} and D.~{Hooper}, \emph{{A History of Dark Matter}}, {\emph{ArXiv
  e-prints} (May, 2016) }, [\href{http://arxiv.org/abs/1605.04909}{{\tt
  1605.04909}}].

\bibitem{Dodelson:93}
S.~{Dodelson} and L.~M. {Widrow}, \emph{{Sterile neutrinos as dark matter}},
  \href{http://dx.doi.org/10.1103/PhysRevLett.72.17}{\emph{Physical Review
  Letters} {\bf 72} (Jan., 1994) 17--20},
  [\href{http://arxiv.org/abs/hep-ph/9303287}{{\tt hep-ph/9303287}}].

\bibitem{Shi:98}
X.~{Shi} and G.~M. {Fuller}, \emph{{New Dark Matter Candidate: Nonthermal
  Sterile Neutrinos}},
  \href{http://dx.doi.org/10.1103/PhysRevLett.82.2832}{\emph{Physical Review
  Letters} {\bf 82} (Apr., 1999) 2832--2835},
  [\href{http://arxiv.org/abs/astro-ph/9810076}{{\tt astro-ph/9810076}}].

\bibitem{Adhikari:16}
R.~{Adhikari}, M.~{Agostini}, N.~A. {Ky}, T.~{Araki}, M.~{Archidiacono},
  M.~{Bahr} et~al., \emph{{A White Paper on keV Sterile Neutrino Dark Matter}},
  {\emph{ArXiv e-prints} (Feb., 2016) },
  [\href{http://arxiv.org/abs/1602.04816}{{\tt 1602.04816}}].

\bibitem{Asaka:05a}
T.~{Asaka}, S.~{Blanchet} and M.~{Shaposhnikov}, \emph{{The {$\nu$}MSM, dark
  matter and neutrino masses [rapid communication]}},
  \href{http://dx.doi.org/10.1016/j.physletb.2005.09.070}{\emph{Physics Letters
  B} {\bf 631} (Dec., 2005) 151--156},
  [\href{http://arxiv.org/abs/hep-ph/0503065}{{\tt hep-ph/0503065}}].

\bibitem{Asaka:05b}
T.~{Asaka} and M.~{Shaposhnikov}, \emph{{The @nMSM, dark matter and baryon
  asymmetry of the universe [rapid communication]}},
  \href{http://dx.doi.org/10.1016/j.physletb.2005.06.020}{\emph{Physics Letters
  B} {\bf 620} (July, 2005) 17--26},
  [\href{http://arxiv.org/abs/hep-ph/0505013}{{\tt hep-ph/0505013}}].

\bibitem{Boyarsky:09a}
A.~{Boyarsky}, O.~{Ruchayskiy} and M.~{Shaposhnikov}, \emph{{The Role of
  Sterile Neutrinos in Cosmology and Astrophysics}},
  \href{http://dx.doi.org/10.1146/annurev.nucl.010909.083654}{\emph{Annual
  Review of Nuclear and Particle Science} {\bf 59} (Nov., 2009) 191--214},
  [\href{http://arxiv.org/abs/0901.0011}{{\tt 0901.0011}}].

\bibitem{Shaposhnikov:08}
M.~{Shaposhnikov}, \emph{{The {$\nu$}MSM, leptonic asymmetries, and properties
  of singlet fermions}},
  \href{http://dx.doi.org/10.1088/1126-6708/2008/08/008}{\emph{Journal of High
  Energy Physics} {\bf 8} (Aug., 2008) 008},
  [\href{http://arxiv.org/abs/0804.4542}{{\tt 0804.4542}}].

\bibitem{Laine:08}
M.~{Laine} and M.~{Shaposhnikov}, \emph{{Sterile neutrino dark matter as a
  consequence of {$\nu$}MSM-induced lepton asymmetry}},
  \href{http://dx.doi.org/10.1088/1475-7516/2008/06/031}{\emph{\jcap} {\bf 6}
  (June, 2008) 031}, [\href{http://arxiv.org/abs/0804.4543}{{\tt 0804.4543}}].

\bibitem{Canetti:12a}
L.~{Canetti}, M.~{Drewes} and M.~{Shaposhnikov}, \emph{{Sterile Neutrinos as
  the Origin of Dark and Baryonic Matter}},
  \href{http://dx.doi.org/10.1103/PhysRevLett.110.061801}{\emph{Physical Review
  Letters} {\bf 110} (Feb., 2013) 061801},
  [\href{http://arxiv.org/abs/1204.3902}{{\tt 1204.3902}}].

\bibitem{Canetti:12b}
L.~{Canetti}, M.~{Drewes}, T.~{Frossard} and M.~{Shaposhnikov}, \emph{{Dark
  matter, baryogenesis and neutrino oscillations from right-handed neutrinos}},
  \href{http://dx.doi.org/10.1103/PhysRevD.87.093006}{\emph{\prd} {\bf 87}
  (May, 2013) 093006}, [\href{http://arxiv.org/abs/1208.4607}{{\tt
  1208.4607}}].

\bibitem{Boyarsky:07a}
A.~{Boyarsky}, D.~{Iakubovskyi}, O.~{Ruchayskiy} and V.~{Savchenko},
  \emph{{Constraints on decaying dark matter from XMM-Newton observations of
  M31}},
  \href{http://dx.doi.org/10.1111/j.1365-2966.2008.13266.x}{\emph{\mnras} {\bf
  387} (July, 2008) 1361--1373}, [\href{http://arxiv.org/abs/0709.2301}{{\tt
  0709.2301}}].

\bibitem{Bulbul:14a}
E.~{Bulbul}, M.~{Markevitch}, A.~{Foster}, R.~K. {Smith}, M.~{Loewenstein} and
  S.~W. {Randall}, \emph{{Detection of an Unidentified Emission Line in the
  Stacked X-Ray Spectrum of Galaxy Clusters}},
  \href{http://dx.doi.org/10.1088/0004-637X/789/1/13}{\emph{\apj} {\bf 789}
  (July, 2014) 13}, [\href{http://arxiv.org/abs/1402.2301}{{\tt 1402.2301}}].

\bibitem{Boyarsky:14a}
A.~{Boyarsky}, O.~{Ruchayskiy}, D.~{Iakubovskyi} and J.~{Franse},
  \emph{{Unidentified Line in X-Ray Spectra of the Andromeda Galaxy and Perseus
  Galaxy Cluster}},
  \href{http://dx.doi.org/10.1103/PhysRevLett.113.251301}{\emph{Physical Review
  Letters} {\bf 113} (Dec., 2014) 251301},
  [\href{http://arxiv.org/abs/1402.4119}{{\tt 1402.4119}}].

\bibitem{Boyarsky:14b}
A.~{Boyarsky}, J.~{Franse}, D.~{Iakubovskyi} and O.~{Ruchayskiy},
  \emph{{Checking the Dark Matter Origin of a 3.53 keV Line with the Milky Way
  Center}},
  \href{http://dx.doi.org/10.1103/PhysRevLett.115.161301}{\emph{Physical Review
  Letters} {\bf 115} (Oct., 2015) 161301},
  [\href{http://arxiv.org/abs/1408.2503}{{\tt 1408.2503}}].

\bibitem{Urban:14}
O.~{Urban}, N.~{Werner}, S.~W. {Allen}, A.~{Simionescu}, J.~S. {Kaastra} and
  L.~E. {Strigari}, \emph{{A Suzaku search for dark matter emission lines in
  the X-ray brightest galaxy clusters}},
  \href{http://dx.doi.org/10.1093/mnras/stv1142}{\emph{\mnras} {\bf 451} (Aug.,
  2015) 2447--2461}, [\href{http://arxiv.org/abs/1411.0050}{{\tt 1411.0050}}].

\bibitem{Iakubovskyi:15b}
D.~{Iakubovskyi}, E.~{Bulbul}, A.~R. {Foster}, D.~{Savchenko} and V.~{Sadova},
  \emph{{Testing the origin of \~{}3.55 keV line in individual galaxy clusters
  observed with XMM-Newton}}, {\emph{ArXiv e-prints} (Aug., 2015) },
  [\href{http://arxiv.org/abs/1508.05186}{{\tt 1508.05186}}].

\bibitem{Iakubovskyi:15c}
D.~{Iakubovskyi}, \emph{{Observation of the new line at \~{}3.55 keV in X-ray
  spectra of galaxies and galaxy clusters}}, {\emph{ArXiv e-prints} (Oct.,
  2015) }, [\href{http://arxiv.org/abs/1510.00358}{{\tt 1510.00358}}].

\bibitem{Ruchayskiy:15}
O.~{Ruchayskiy}, A.~{Boyarsky}, D.~{Iakubovskyi}, E.~{Bulbul}, D.~{Eckert},
  J.~{Franse} et~al., \emph{{Searching for decaying dark matter in deep
  XMM-Newton observation of the Draco dwarf spheroidal}}, {\emph{ArXiv
  e-prints} (Dec., 2015) }, [\href{http://arxiv.org/abs/1512.07217}{{\tt
  1512.07217}}].

\bibitem{Abazajian:14}
K.~N. {Abazajian}, \emph{{Resonantly Produced 7 keV Sterile Neutrino Dark
  Matter Models and the Properties of Milky Way Satellites}},
  \href{http://dx.doi.org/10.1103/PhysRevLett.112.161303}{\emph{Physical Review
  Letters} {\bf 112} (Apr., 2014) 161303},
  [\href{http://arxiv.org/abs/1403.0954}{{\tt 1403.0954}}].

\bibitem{Venumadhav:15}
T.~{Venumadhav}, F.-Y. {Cyr-Racine}, K.~N. {Abazajian} and C.~M. {Hirata},
  \emph{{Sterile neutrino dark matter: A tale of weak interactions in the
  strong coupling epoch}}, {\emph{ArXiv e-prints} (July, 2015) },
  [\href{http://arxiv.org/abs/1507.06655}{{\tt 1507.06655}}].

\bibitem{Lovell:15}
M.~R. {Lovell}, S.~{Bose}, A.~{Boyarsky}, S.~{Cole}, C.~S. {Frenk},
  V.~{Gonzalez-Perez} et~al., \emph{{Satellite galaxies in semi-analytic models
  of galaxy formation with sterile neutrino dark matter}}, {\emph{ArXiv
  e-prints} (Nov., 2015) }, [\href{http://arxiv.org/abs/1511.04078}{{\tt
  1511.04078}}].

\bibitem{Horiuchi:15}
S.~{Horiuchi}, B.~{Bozek}, K.~N. {Abazajian}, M.~{Boylan-Kolchin}, J.~S.
  {Bullock}, S.~{Garrison-Kimmel} et~al., \emph{{Properties of resonantly
  produced sterile neutrino dark matter subhaloes}},
  \href{http://dx.doi.org/10.1093/mnras/stv2922}{\emph{\mnras} {\bf 456} (Mar.,
  2016) 4346--4353}, [\href{http://arxiv.org/abs/1512.04548}{{\tt
  1512.04548}}].

\bibitem{Bose:16a}
S.~{Bose}, W.~A. {Hellwing}, C.~S. {Frenk}, A.~{Jenkins}, M.~R. {Lovell}, J.~C.
  {Helly} et~al., \emph{{Substructure and galaxy formation in the Copernicus
  Complexio warm dark matter simulations}}, {\emph{ArXiv e-prints} (Apr., 2016)
  }, [\href{http://arxiv.org/abs/1604.07409}{{\tt 1604.07409}}].

\bibitem{Schneider:16}
A.~{Schneider}, \emph{{Astrophysical constraints on resonantly produced sterile
  neutrino dark matter}}, {\emph{ArXiv e-prints} (Jan., 2016) },
  [\href{http://arxiv.org/abs/1601.07553}{{\tt 1601.07553}}].

\bibitem{Bozek:15}
B.~{Bozek}, M.~{Boylan-Kolchin}, S.~{Horiuchi}, S.~{Garrison-Kimmel},
  K.~{Abazajian} and J.~S. {Bullock}, \emph{{Resonant Sterile Neutrino Dark
  Matter in the Local and High-z Universe}}, {\emph{ArXiv e-prints} (Dec.,
  2015) }, [\href{http://arxiv.org/abs/1512.04544}{{\tt 1512.04544}}].

\bibitem{Bose:15}
S.~{Bose}, W.~A. {Hellwing}, C.~S. {Frenk}, A.~{Jenkins}, M.~R. {Lovell}, J.~C.
  {Helly} et~al., \emph{{The Copernicus Complexio: statistical properties of
  warm dark matter haloes}},
  \href{http://dx.doi.org/10.1093/mnras/stv2294}{\emph{\mnras} {\bf 455} (Jan.,
  2016) 318--333}, [\href{http://arxiv.org/abs/1507.01998}{{\tt 1507.01998}}].

\bibitem{Schneider:14}
A.~{Schneider}, \emph{{Structure formation with suppressed small-scale
  perturbations}}, \href{http://dx.doi.org/10.1093/mnras/stv1169}{\emph{\mnras}
  {\bf 451} (Aug., 2015) 3117--3130},
  [\href{http://arxiv.org/abs/1412.2133}{{\tt 1412.2133}}].

\bibitem{Li:15}
R.~{Li}, C.~S. {Frenk}, S.~{Cole}, L.~{Gao}, S.~{Bose} and W.~A. {Hellwing},
  \emph{{Constraints on the identity of the dark matter from strong
  gravitational lenses}},
  \href{http://dx.doi.org/10.1093/mnras/stw939}{\emph{\mnras} (Apr., 2016) },
  [\href{http://arxiv.org/abs/1512.06507}{{\tt 1512.06507}}].

\bibitem{Ludlow:16}
A.~D. {Ludlow}, S.~{Bose}, R.~E. {Angulo}, L.~{Wang}, W.~A. {Hellwing}, J.~F.
  {Navarro} et~al., \emph{{The Mass-Concentration-Redshift Relation of Cold and
  Warm Dark Matter Halos}}, {\emph{ArXiv e-prints} (Jan., 2016) },
  [\href{http://arxiv.org/abs/1601.02624}{{\tt 1601.02624}}].

\bibitem{Wang:15}
M.-Y. {Wang}, L.~E. {Strigari}, M.~R. {Lovell}, C.~S. {Frenk} and A.~R.
  {Zentner}, \emph{{Mass assembly history and infall time of the Fornax dwarf
  spheroidal galaxy}},
  \href{http://dx.doi.org/10.1093/mnras/stw220}{\emph{\mnras} {\bf 457} (Apr.,
  2016) 4248--4261}, [\href{http://arxiv.org/abs/1509.04308}{{\tt
  1509.04308}}].

\bibitem{Kamada:16}
A.~{Kamada}, K.~T. {Inoue} and T.~{Takahashi}, \emph{{Constraints on mixed dark
  matter from anomalous strong lens systems}}, {\emph{ArXiv e-prints} (Apr.,
  2016) }, [\href{http://arxiv.org/abs/1604.01489}{{\tt 1604.01489}}].

\bibitem{Mahdi:16}
H.~S. {Mahdi}, P.~J. {Elahi}, G.~F. {Lewis} and C.~{Power}, \emph{{Matter in
  the beam: Weak lensing, substructures and the temperature of dark matter}},
  {\emph{ArXiv e-prints} (May, 2016) },
  [\href{http://arxiv.org/abs/1605.08428}{{\tt 1605.08428}}].

\bibitem{Barkana:00}
R.~{Barkana} and A.~{Loeb}, \emph{{In the beginning: the first sources of light
  and the reionization of the universe}},
  \href{http://dx.doi.org/10.1016/S0370-1573(01)00019-9}{\emph{\physrep} {\bf
  349} (July, 2001) 125--238},
  [\href{http://arxiv.org/abs/astro-ph/0010468}{{\tt astro-ph/0010468}}].

\bibitem{Fan:06}
X.~{Fan}, C.~L. {Carilli} and B.~{Keating}, \emph{{Observational Constraints on
  Cosmic Reionization}},
  \href{http://dx.doi.org/10.1146/annurev.astro.44.051905.092514}{\emph{\araa}
  {\bf 44} (Sept., 2006) 415--462},
  [\href{http://arxiv.org/abs/astro-ph/0602375}{{\tt astro-ph/0602375}}].

\bibitem{Natarajan:14}
A.~{Natarajan} and N.~{Yoshida}, \emph{{The Dark Ages of the Universe and
  hydrogen reionization}},
  \href{http://dx.doi.org/10.1093/ptep/ptu067}{\emph{Progress of Theoretical
  and Experimental Physics} {\bf 2014} (June, 2014) 06B112},
  [\href{http://arxiv.org/abs/1404.7146}{{\tt 1404.7146}}].

\bibitem{Ferrara:14}
A.~{Ferrara} and S.~{Pandolfi}, \emph{{Reionization of the Intergalactic
  Medium}}, {\emph{ArXiv e-prints} (Sept., 2014) },
  [\href{http://arxiv.org/abs/1409.4946}{{\tt 1409.4946}}].

\bibitem{McQuinn:15}
M.~{McQuinn}, \emph{{The Evolution of the Intergalactic Medium}}, {\emph{ArXiv
  e-prints} (Nov., 2015) }, [\href{http://arxiv.org/abs/1512.00086}{{\tt
  1512.00086}}].

\bibitem{Mesinger:16}
A.~{Mesinger}, ed., \emph{{Understanding the Epoch of Cosmic Reionization}},
  vol.~423 of \emph{Astrophysics and Space Science Library}, 2016.
\newblock 10.1007/978-3-319-21957-8.

\bibitem{Dayal:14}
P.~{Dayal}, A.~{Mesinger} and F.~{Pacucci}, \emph{{Early Galaxy Formation in
  Warm Dark Matter Cosmologies}},
  \href{http://dx.doi.org/10.1088/0004-637X/806/1/67}{\emph{\apj} {\bf 806}
  (June, 2015) 67}, [\href{http://arxiv.org/abs/1408.1102}{{\tt 1408.1102}}].

\bibitem{Haiman:00}
Z.~{Haiman}, T.~{Abel} and P.~{Madau}, \emph{{Photon Consumption in Minihalos
  during Cosmological Reionization}},
  \href{http://dx.doi.org/10.1086/320232}{\emph{\apj} {\bf 551} (Apr., 2001)
  599--607}, [\href{http://arxiv.org/abs/astro-ph/0009125}{{\tt
  astro-ph/0009125}}].

\bibitem{Barkana:02}
R.~{Barkana} and A.~{Loeb}, \emph{{Effective Screening Due to Minihalos during
  the Epoch of Reionization}},
  \href{http://dx.doi.org/10.1086/342313}{\emph{\apj} {\bf 578} (Oct., 2002)
  1--11}, [\href{http://arxiv.org/abs/astro-ph/0204139}{{\tt
  astro-ph/0204139}}].

\bibitem{Shapiro:03}
P.~R. {Shapiro}, I.~T. {Iliev} and A.~C. {Raga}, \emph{{Photoevaporation of
  cosmological minihaloes during reionization}},
  \href{http://dx.doi.org/10.1111/j.1365-2966.2004.07364.x}{\emph{\mnras} {\bf
  348} (Mar., 2004) 753--782},
  [\href{http://arxiv.org/abs/astro-ph/0307266}{{\tt astro-ph/0307266}}].

\bibitem{Iliev:04a}
I.~T. {Iliev}, P.~R. {Shapiro} and A.~C. {Raga}, \emph{{Minihalo
  photoevaporation during cosmic reionization: evaporation times and photon
  consumption rates}},
  \href{http://dx.doi.org/10.1111/j.1365-2966.2005.09155.x}{\emph{\mnras} {\bf
  361} (Aug., 2005) 405--414},
  [\href{http://arxiv.org/abs/astro-ph/0408408}{{\tt astro-ph/0408408}}].

\bibitem{Ciardi:05}
B.~{Ciardi}, E.~{Scannapieco}, F.~{Stoehr}, A.~{Ferrara}, I.~T. {Iliev} and
  P.~R. {Shapiro}, \emph{{The effect of minihaloes on cosmic reionization}},
  \href{http://dx.doi.org/10.1111/j.1365-2966.2005.09908.x}{\emph{\mnras} {\bf
  366} (Feb., 2006) 689--696},
  [\href{http://arxiv.org/abs/astro-ph/0511623}{{\tt astro-ph/0511623}}].

\bibitem{McQuinn:06}
M.~{McQuinn}, A.~{Lidz}, O.~{Zahn}, S.~{Dutta}, L.~{Hernquist} and
  M.~{Zaldarriaga}, \emph{{The morphology of HII regions during reionization}},
  \href{http://dx.doi.org/10.1111/j.1365-2966.2007.11489.x}{\emph{\mnras} {\bf
  377} (May, 2007) 1043--1063},
  [\href{http://arxiv.org/abs/astro-ph/0610094}{{\tt astro-ph/0610094}}].

\bibitem{Alvarez:10}
M.~A. {Alvarez} and T.~{Abel}, \emph{{The Effect of Absorption Systems on
  Cosmic Reionization}},
  \href{http://dx.doi.org/10.1088/0004-637X/747/2/126}{\emph{\apj} {\bf 747}
  (Mar., 2012) 126}, [\href{http://arxiv.org/abs/1003.6132}{{\tt 1003.6132}}].

\bibitem{Park:16}
H.~{Park}, P.~R. {Shapiro}, J.-h. {Choi}, N.~{Yoshida}, S.~{Hirano} and
  K.~{Ahn}, \emph{{The Hydrodynamic Feedback of Cosmic Reionization on
  Small-Scale Structures and Its Impact on Photon Consumption during the Epoch
  of Reionization}}, {\emph{ArXiv e-prints} (Feb., 2016) },
  [\href{http://arxiv.org/abs/1602.06472}{{\tt 1602.06472}}].

\bibitem{Yue:12}
B.~{Yue} and X.~{Chen}, \emph{{Reionization in the Warm Dark Matter Model}},
  \href{http://dx.doi.org/10.1088/0004-637X/747/2/127}{\emph{\apj} {\bf 747}
  (Mar., 2012) 127}, [\href{http://arxiv.org/abs/1201.3686}{{\tt 1201.3686}}].

\bibitem{Barkana:01}
R.~{Barkana}, Z.~{Haiman} and J.~P. {Ostriker}, \emph{{Constraints on Warm Dark
  Matter from Cosmological Reionization}},
  \href{http://dx.doi.org/10.1086/322393}{\emph{\apj} {\bf 558} (Sept., 2001)
  482--496}, [\href{http://arxiv.org/abs/astro-ph/0102304}{{\tt
  astro-ph/0102304}}].

\bibitem{Somerville:03}
R.~S. {Somerville}, J.~S. {Bullock} and M.~{Livio}, \emph{{The Epoch of
  Reionization in Models with Reduced Small-Scale Power}},
  \href{http://dx.doi.org/10.1086/376686}{\emph{\apj} {\bf 593} (Aug., 2003)
  616--621}, [\href{http://arxiv.org/abs/astro-ph/0303481}{{\tt
  astro-ph/0303481}}].

\bibitem{Yoshida:03}
N.~{Yoshida}, A.~{Sokasian}, L.~{Hernquist} and V.~{Springel}, \emph{{Early
  Structure Formation and Reionization in a Warm Dark Matter Cosmology}},
  \href{http://dx.doi.org/10.1086/376963}{\emph{\apjl} {\bf 591} (July, 2003)
  L1--L4}, [\href{http://arxiv.org/abs/arXiv:astro-ph/0303622}{{\tt
  arXiv:astro-ph/0303622}}].

\bibitem{Schultz:14}
C.~{Schultz}, J.~{O{\~n}orbe}, K.~N. {Abazajian} and J.~S. {Bullock},
  \emph{{The high-z universe confronts warm dark matter: Galaxy counts,
  reionization and the nature of dark matter}},
  \href{http://dx.doi.org/10.1093/mnras/stu976}{\emph{\mnras} {\bf 442} (Aug.,
  2014) 1597--1609}, [\href{http://arxiv.org/abs/1401.3769}{{\tt 1401.3769}}].

\bibitem{Dayal:15}
P.~{Dayal}, T.~R. {Choudhury}, V.~{Bromm} and F.~{Pacucci}, \emph{{Reionizing
  the Universe in Warm Dark Matter cosmologies}}, {\emph{ArXiv e-prints} (Jan.,
  2015) }, [\href{http://arxiv.org/abs/1501.02823}{{\tt 1501.02823}}].

\bibitem{Furlanetto:04a}
S.~R. {Furlanetto}, M.~{Zaldarriaga} and L.~{Hernquist}, \emph{{The Growth of H
  II Regions During Reionization}},
  \href{http://dx.doi.org/10.1086/423025}{\emph{\apj} {\bf 613} (Sept., 2004)
  1--15}, [\href{http://arxiv.org/abs/astro-ph/0403697}{{\tt
  astro-ph/0403697}}].

\bibitem{Inoue:06}
A.~K. {Inoue}, I.~{Iwata} and J.-M. {Deharveng}, \emph{{The escape fraction of
  ionizing photons from galaxies at z = 0-6}},
  \href{http://dx.doi.org/10.1111/j.1745-3933.2006.00195.x}{\emph{\mnras} {\bf
  371} (Sept., 2006) L1--L5},
  [\href{http://arxiv.org/abs/astro-ph/0605526}{{\tt astro-ph/0605526}}].

\bibitem{Razoumov:06}
A.~O. {Razoumov} and J.~{Sommer-Larsen}, \emph{{Escape of Ionizing Radiation
  from Star-forming Regions in Young Galaxies}},
  \href{http://dx.doi.org/10.1086/509636}{\emph{\apjl} {\bf 651} (Nov., 2006)
  L89--L92}, [\href{http://arxiv.org/abs/astro-ph/0609545}{{\tt
  astro-ph/0609545}}].

\bibitem{Faisst:16}
A.~L. {Faisst}, \emph{{Revisiting the LyC Escape Fraction Crisis: Predictions
  for z$>$6 from Locals}}, {\emph{ArXiv e-prints} (May, 2016) },
  [\href{http://arxiv.org/abs/1605.06507}{{\tt 1605.06507}}].

\bibitem{Mitra:15}
S.~{Mitra}, T.~R. {Choudhury} and A.~{Ferrara}, \emph{{Cosmic reionization
  after Planck}}, \href{http://dx.doi.org/10.1093/mnrasl/slv134}{\emph{\mnras}
  {\bf 454} (Nov., 2015) L76--L80},
  [\href{http://arxiv.org/abs/1505.05507}{{\tt 1505.05507}}].

\bibitem{Iliev:04b}
I.~T. {Iliev}, E.~{Scannapieco} and P.~R. {Shapiro}, \emph{{The Impact of
  Small-Scale Structure on Cosmological Ionization Fronts and Reionization}},
  \href{http://dx.doi.org/10.1086/429083}{\emph{\apj} {\bf 624} (May, 2005)
  491--504}, [\href{http://arxiv.org/abs/astro-ph/0411035}{{\tt
  astro-ph/0411035}}].

\bibitem{Leitherer:99}
C.~{Leitherer}, D.~{Schaerer}, J.~D. {Goldader}, R.~M.~G. {Delgado},
  C.~{Robert}, D.~F. {Kune} et~al., \emph{{Starburst99: Synthesis Models for
  Galaxies with Active Star Formation}},
  \href{http://dx.doi.org/10.1086/313233}{\emph{\apjs} {\bf 123} (July, 1999)
  3--40}, [\href{http://arxiv.org/abs/astro-ph/9902334}{{\tt
  astro-ph/9902334}}].

\bibitem{Becker:15}
G.~D. {Becker}, J.~S. {Bolton} and A.~{Lidz}, \emph{{Reionisation and
  High-Redshift Galaxies: The View from Quasar Absorption Lines}},
  \href{http://dx.doi.org/10.1017/pasa.2015.45}{\emph{\pasa} {\bf 32} (Dec.,
  2015) e045}, [\href{http://arxiv.org/abs/1510.03368}{{\tt 1510.03368}}].

\bibitem{Wyithe:09}
J.~S.~B. {Wyithe}, A.~M. {Hopkins}, M.~D. {Kistler}, H.~{Y{\"u}ksel} and J.~F.
  {Beacom}, \emph{{Determining the escape fraction of ionizing photons during
  reionization with the GRB-derived star formation rate}},
  \href{http://dx.doi.org/10.1111/j.1365-2966.2009.15834.x}{\emph{\mnras} {\bf
  401} (Feb., 2010) 2561--2571}, [\href{http://arxiv.org/abs/0908.0193}{{\tt
  0908.0193}}].

\bibitem{Behroozi:12}
P.~S. {Behroozi}, R.~H. {Wechsler} and C.~{Conroy}, \emph{{The Average Star
  Formation Histories of Galaxies in Dark Matter Halos from z = 0-8}},
  \href{http://dx.doi.org/10.1088/0004-637X/770/1/57}{\emph{\apj} {\bf 770}
  (June, 2013) 57}, [\href{http://arxiv.org/abs/1207.6105}{{\tt 1207.6105}}].

\bibitem{Wise:14}
J.~H. {Wise}, V.~G. {Demchenko}, M.~T. {Halicek}, M.~L. {Norman}, M.~J. {Turk},
  T.~{Abel} et~al., \emph{{The birth of a galaxy - III. Propelling reionization
  with the faintest galaxies}},
  \href{http://dx.doi.org/10.1093/mnras/stu979}{\emph{\mnras} {\bf 442} (Aug.,
  2014) 2560--2579}, [\href{http://arxiv.org/abs/1403.6123}{{\tt 1403.6123}}].

\bibitem{Siana:15}
B.~{Siana}, A.~E. {Shapley}, K.~R. {Kulas}, D.~B. {Nestor}, C.~C. {Steidel},
  H.~I. {Teplitz} et~al., \emph{{A Deep Hubble Space Telescope and Keck Search
  for Definitive Identification of Lyman Continuum Emitters at z\~{}3.1}},
  \href{http://dx.doi.org/10.1088/0004-637X/804/1/17}{\emph{\apj} {\bf 804}
  (May, 2015) 17}, [\href{http://arxiv.org/abs/1502.06978}{{\tt 1502.06978}}].

\bibitem{Grazian:15}
A.~{Grazian}, E.~{Giallongo}, R.~{Gerbasi}, F.~{Fiore}, A.~{Fontana}, O.~{Le
  F{\`e}vre} et~al., \emph{{The Lyman continuum escape fraction of galaxies at
  z = 3.3 in the VUDS-LBC/COSMOS field}},
  \href{http://dx.doi.org/10.1051/0004-6361/201526396}{\emph{\aap} {\bf 585}
  (Jan., 2016) A48}, [\href{http://arxiv.org/abs/1509.01101}{{\tt
  1509.01101}}].

\bibitem{Izotov:16}
Y.~I. {Izotov}, I.~{Orlitov{\'a}}, D.~{Schaerer}, T.~X. {Thuan}, A.~{Verhamme},
  N.~G. {Guseva} et~al., \emph{{Eight per cent leakage of Lyman continuum
  photons from a compact, star-forming dwarf galaxy}},
  \href{http://dx.doi.org/10.1038/nature16456}{\emph{\nat} {\bf 529} (Jan.,
  2016) 178--180}.

\bibitem{Borthakur:14}
S.~{Borthakur}, T.~M. {Heckman}, C.~{Leitherer} and R.~A. {Overzier}, \emph{{A
  local clue to the reionization of the universe}},
  \href{http://dx.doi.org/10.1126/science.1254214}{\emph{Science} {\bf 346}
  (Oct., 2014) 216--219}, [\href{http://arxiv.org/abs/1410.3511}{{\tt
  1410.3511}}].

\bibitem{Ma:16}
X.~{Ma}, P.~F. {Hopkins}, D.~{Kasen}, E.~{Quataert}, C.-A. {Faucher-Giguere},
  D.~{Keres} et~al., \emph{{Binary Stars Can Provide the ''Missing Photons''
  Needed for Reionization}}, {\emph{ArXiv e-prints} (Jan., 2016) },
  [\href{http://arxiv.org/abs/1601.07559}{{\tt 1601.07559}}].

\bibitem{Press:74}
W.~H. {Press} and P.~{Schechter}, \emph{{Formation of Galaxies and Clusters of
  Galaxies by Self-Similar Gravitational Condensation}},
  \href{http://dx.doi.org/10.1086/152650}{\emph{\apj} {\bf 187} (Feb., 1974)
  425--438}.

\bibitem{Bond:91}
J.~R. {Bond}, S.~{Cole}, G.~{Efstathiou} and N.~{Kaiser}, \emph{{Excursion set
  mass functions for hierarchical Gaussian fluctuations}},
  \href{http://dx.doi.org/10.1086/170520}{\emph{\apj} {\bf 379} (Oct., 1991)
  440--460}.

\bibitem{Bower:91}
R.~G. {Bower}, \emph{{The evolution of groups of galaxies in the
  Press-Schechter formalism}},
  \href{http://dx.doi.org/10.1093/mnras/248.2.332}{\emph{\mnras} {\bf 248}
  (Jan., 1991) 332--352}.

\bibitem{Lacey:93}
C.~{Lacey} and S.~{Cole}, \emph{{Merger rates in hierarchical models of galaxy
  formation}}, \href{http://dx.doi.org/10.1093/mnras/262.3.627}{\emph{\mnras}
  {\bf 262} (June, 1993) 627--649}.

\bibitem{Planck2015-XIII}
{Planck Collaboration}, P.~A.~R. {Ade}, N.~{Aghanim}, M.~{Arnaud},
  M.~{Ashdown}, J.~{Aumont} et~al., \emph{{Planck 2015 results. XIII.
  Cosmological parameters}}, {\emph{ArXiv e-prints} (Feb., 2015) },
  [\href{http://arxiv.org/abs/1502.01589}{{\tt 1502.01589}}].

\bibitem{Eke:96}
V.~R. {Eke}, S.~{Cole} and C.~S. {Frenk}, \emph{{Cluster evolution as a
  diagnostic for Omega}},
  \href{http://dx.doi.org/10.1093/mnras/282.1.263}{\emph{\mnras} {\bf 282}
  (Sept., 1996) }, [\href{http://arxiv.org/abs/astro-ph/9601088}{{\tt
  astro-ph/9601088}}].

\bibitem{Heath:77}
D.~J. {Heath}, \emph{{The growth of density perturbations in zero pressure
  Friedmann-Lemaitre universes}},
  \href{http://dx.doi.org/10.1093/mnras/179.3.351}{\emph{\mnras} {\bf 179}
  (May, 1977) 351--358}.

\bibitem{Eisenstein:97a}
D.~J. {Eisenstein}, \emph{{An Analytic Expression for the Growth Function in a
  Flat Universe with a Cosmological Constant}}, {\emph{ArXiv Astrophysics
  e-prints} (Sept., 1997) }, [\href{http://arxiv.org/abs/astro-ph/9709054}{{\tt
  astro-ph/9709054}}].

\bibitem{Ghiglieri:15}
J.~{Ghiglieri} and M.~{Laine}, \emph{{Improved determination of sterile
  neutrino dark matter spectrum}},
  \href{http://dx.doi.org/10.1007/JHEP11(2015)171}{\emph{Journal of High Energy
  Physics} {\bf 11} (Nov., 2015) 171},
  [\href{http://arxiv.org/abs/1506.06752}{{\tt 1506.06752}}].

\bibitem{Bertschinger:06}
E.~{Bertschinger}, \emph{{Effects of cold dark matter decoupling and pair
  annihilation on cosmological perturbations}},
  \href{http://dx.doi.org/10.1103/PhysRevD.74.063509}{\emph{\prd} {\bf 74}
  (Sept., 2006) 063509}, [\href{http://arxiv.org/abs/astro-ph/0607319}{{\tt
  astro-ph/0607319}}].

\bibitem{Benson:12}
A.~J. {Benson}, A.~{Farahi}, S.~{Cole}, L.~A. {Moustakas}, A.~{Jenkins},
  M.~{Lovell} et~al., \emph{{Dark matter halo merger histories beyond cold dark
  matter - I. Methods and application to warm dark matter}},
  \href{http://dx.doi.org/10.1093/mnras/sts159}{\emph{\mnras} {\bf 428} (Jan.,
  2013) 1774--1789}, [\href{http://arxiv.org/abs/1209.3018}{{\tt 1209.3018}}].

\bibitem{Schneider:13}
A.~{Schneider}, R.~E. {Smith} and D.~{Reed}, \emph{{Halo mass function and the
  free streaming scale}},
  \href{http://dx.doi.org/10.1093/mnras/stt829}{\emph{\mnras} {\bf 433} (Aug.,
  2013) 1573--1587}, [\href{http://arxiv.org/abs/1303.0839}{{\tt 1303.0839}}].

\bibitem{Haiman:96}
Z.~{Haiman}, M.~J. {Rees} and A.~{Loeb}, \emph{{Destruction of Molecular
  Hydrogen during Cosmological Reionization}}, {\emph{\apj} {\bf 476} (Feb.,
  1997) 458--463}, [\href{http://arxiv.org/abs/astro-ph/9608130}{{\tt
  astro-ph/9608130}}].

\bibitem{Loeb:10book}
A.~{Loeb}, \emph{{How Did the First Stars and Galaxies Form?}}
\newblock 2010.

\bibitem{Finlator:12}
K.~{Finlator}, S.~P. {Oh}, F.~{{\"O}zel} and R.~{Dav{\'e}}, \emph{{Gas clumping
  in self-consistent reionization models}},
  \href{http://dx.doi.org/10.1111/j.1365-2966.2012.22114.x}{\emph{\mnras} {\bf
  427} (Dec., 2012) 2464--2479}.

\bibitem{Sheth:98}
R.~K. {Sheth}, \emph{{An excursion set model for the distribution of dark
  matter and dark matter haloes}},
  \href{http://dx.doi.org/10.1046/j.1365-8711.1998.01976.x}{\emph{\mnras} {\bf
  300} (Nov., 1998) 1057--1070},
  [\href{http://arxiv.org/abs/astro-ph/9805319}{{\tt astro-ph/9805319}}].

\bibitem{Furlanetto:15b}
S.~R. {Furlanetto} and S.~P. {Oh}, \emph{{Reionization through the lens of
  percolation theory}},
  \href{http://dx.doi.org/10.1093/mnras/stw104}{\emph{\mnras} {\bf 457} (Apr.,
  2016) 1813--1827}, [\href{http://arxiv.org/abs/1511.01521}{{\tt
  1511.01521}}].

\bibitem{Kuhlen:12}
M.~{Kuhlen} and C.-A. {Faucher-Gigu{\`e}re}, \emph{{Concordance models of
  reionization: implications for faint galaxies and escape fraction
  evolution}},
  \href{http://dx.doi.org/10.1111/j.1365-2966.2012.20924.x}{\emph{\mnras} {\bf
  423} (June, 2012) 862--876}, [\href{http://arxiv.org/abs/1201.0757}{{\tt
  1201.0757}}].

\bibitem{Robertson:13}
B.~E. {Robertson}, S.~R. {Furlanetto}, E.~{Schneider}, S.~{Charlot}, R.~S.
  {Ellis}, D.~P. {Stark} et~al., \emph{{New Constraints on Cosmic Reionization
  from the 2012 Hubble Ultra Deep Field Campaign}},
  \href{http://dx.doi.org/10.1088/0004-637X/768/1/71}{\emph{\apj} {\bf 768}
  (May, 2013) 71}, [\href{http://arxiv.org/abs/1301.1228}{{\tt 1301.1228}}].

\bibitem{Robertson:15}
B.~E. {Robertson}, R.~S. {Ellis}, S.~R. {Furlanetto} and J.~S. {Dunlop},
  \emph{{Cosmic Reionization and Early Star-forming Galaxies: A Joint Analysis
  of New Constraints from Planck and the Hubble Space Telescope}},
  \href{http://dx.doi.org/10.1088/2041-8205/802/2/L19}{\emph{\apjl} {\bf 802}
  (Apr., 2015) L19}, [\href{http://arxiv.org/abs/1502.02024}{{\tt
  1502.02024}}].

\bibitem{Fan:05}
X.~{Fan}, M.~A. {Strauss}, R.~H. {Becker}, R.~L. {White}, J.~E. {Gunn}, G.~R.
  {Knapp} et~al., \emph{{Constraining the Evolution of the Ionizing Background
  and the Epoch of Reionization with z\~{}6 Quasars. II. A Sample of 19
  Quasars}}, \href{http://dx.doi.org/10.1086/504836}{\emph{\aj} {\bf 132}
  (July, 2006) 117--136},
  [\href{http://arxiv.org/abs/arXiv:astro-ph/0512082}{{\tt
  arXiv:astro-ph/0512082}}].

\bibitem{McGreer:14}
I.~D. {McGreer}, A.~{Mesinger} and V.~{D'Odorico}, \emph{{Model-independent
  evidence in favour of an end to reionization by z$\sim$6}},
  \href{http://dx.doi.org/10.1093/mnras/stu2449}{\emph{\mnras} {\bf 447} (Feb.,
  2015) 499--505}, [\href{http://arxiv.org/abs/1411.5375}{{\tt 1411.5375}}].

\bibitem{Bolton:11}
J.~S. {Bolton}, M.~G. {Haehnelt}, S.~J. {Warren}, P.~C. {Hewett}, D.~J.
  {Mortlock}, B.~P. {Venemans} et~al., \emph{{How neutral is the intergalactic
  medium surrounding the redshift z = 7.085 quasar ULAS J1120+0641?}},
  \href{http://dx.doi.org/10.1111/j.1745-3933.2011.01100.x}{\emph{\mnras} {\bf
  416} (Sept., 2011) L70--L74}, [\href{http://arxiv.org/abs/1106.6089}{{\tt
  1106.6089}}].

\bibitem{Schroeder:12}
J.~{Schroeder}, A.~{Mesinger} and Z.~{Haiman}, \emph{{Evidence of Gunn-Peterson
  damping wings in high-z quasar spectra: strengthening the case for incomplete
  reionization at z $\sim$ 6-7}},
  \href{http://dx.doi.org/10.1093/mnras/sts253}{\emph{\mnras} {\bf 428} (Feb.,
  2013) 3058--3071}, [\href{http://arxiv.org/abs/1204.2838}{{\tt 1204.2838}}].

\bibitem{Bosman:15}
S.~E.~I. {Bosman} and G.~D. {Becker}, \emph{{Re-examining the case for neutral
  gas near the redshift 7 quasar ULAS J1120+0641}},
  \href{http://dx.doi.org/10.1093/mnras/stv1336}{\emph{\mnras} {\bf 452}
  (Sept., 2015) 1105--1111}, [\href{http://arxiv.org/abs/1505.06880}{{\tt
  1505.06880}}].

\bibitem{Chornock:13}
R.~{Chornock}, E.~{Berger}, D.~B. {Fox}, R.~{Lunnan}, M.~R. {Drout}, W.-f.
  {Fong} et~al., \emph{{GRB 130606A as a Probe of the Intergalactic Medium and
  the Interstellar Medium in a Star-forming Galaxy in the First Gyr after the
  Big Bang}}, \href{http://dx.doi.org/10.1088/0004-637X/774/1/26}{\emph{\apj}
  {\bf 774} (Sept., 2013) 26}, [\href{http://arxiv.org/abs/1306.3949}{{\tt
  1306.3949}}].

\bibitem{Ota:07}
K.~{Ota}, M.~{Iye}, N.~{Kashikawa}, K.~{Shimasaku}, M.~{Kobayashi}, T.~{Totani}
  et~al., \emph{{Reionization and Galaxy Evolution Probed by z = 7 Ly{$\alpha$}
  Emitters}}, \href{http://dx.doi.org/10.1086/529006}{\emph{\apj} {\bf 677}
  (Apr., 2008) 12--26}, [\href{http://arxiv.org/abs/0707.1561}{{\tt
  0707.1561}}].

\bibitem{Ouchi:10}
M.~{Ouchi}, K.~{Shimasaku}, H.~{Furusawa}, T.~{Saito}, M.~{Yoshida},
  M.~{Akiyama} et~al., \emph{{Statistics of 207 Ly{$\alpha$} Emitters at a
  Redshift Near 7: Constraints on Reionization and Galaxy Formation Models}},
  \href{http://dx.doi.org/10.1088/0004-637X/723/1/869}{\emph{\apj} {\bf 723}
  (Nov., 2010) 869--894}, [\href{http://arxiv.org/abs/1007.2961}{{\tt
  1007.2961}}].

\bibitem{Planck2016-XLVII}
{Planck Collaboration}, R.~{Adam}, N.~{Aghanim}, M.~{Ashdown}, J.~{Aumont},
  C.~{Baccigalupi} et~al., \emph{{Planck 2016 intermediate results. XLVII.
  Planck constraints on reionization history}}, {\emph{ArXiv e-prints} (May,
  2016) }, [\href{http://arxiv.org/abs/1605.03507}{{\tt 1605.03507}}].

\bibitem{Planck2016-XLVI}
{Planck Collaboration}, N.~{Aghanim}, M.~{Ashdown}, J.~{Aumont},
  C.~{Baccigalupi}, M.~{Ballardini} et~al., \emph{{Planck intermediate results.
  XLVI. Reduction of large-scale systematic effects in HFI polarization maps
  and estimation of the reionization optical depth}}, {\emph{ArXiv e-prints}
  (May, 2016) }, [\href{http://arxiv.org/abs/1605.02985}{{\tt 1605.02985}}].

\bibitem{Hinshaw:12}
G.~{Hinshaw}, D.~{Larson}, E.~{Komatsu}, D.~N. {Spergel}, C.~L. {Bennett},
  J.~{Dunkley} et~al., \emph{{Nine-year Wilkinson Microwave Anisotropy Probe
  (WMAP) Observations: Cosmological Parameter Results}},
  \href{http://dx.doi.org/10.1088/0067-0049/208/2/19}{\emph{\apjs} {\bf 208}
  (Oct., 2013) 19}, [\href{http://arxiv.org/abs/1212.5226}{{\tt 1212.5226}}].

\bibitem{Madau:15}
P.~{Madau} and F.~{Haardt}, \emph{{Cosmic Reionization after Planck: Could
  Quasars Do It All?}},
  \href{http://dx.doi.org/10.1088/2041-8205/813/1/L8}{\emph{\apjl} {\bf 813}
  (Nov., 2015) L8}, [\href{http://arxiv.org/abs/1507.07678}{{\tt 1507.07678}}].

\bibitem{Dixon:15}
K.~L. {Dixon}, I.~T. {Iliev}, G.~{Mellema}, K.~{Ahn} and P.~R. {Shapiro},
  \emph{{The large-scale observational signatures of low-mass galaxies during
  reionization}}, \href{http://dx.doi.org/10.1093/mnras/stv2887}{\emph{\mnras}
  {\bf 456} (Mar., 2016) 3011--3029},
  [\href{http://arxiv.org/abs/1512.03836}{{\tt 1512.03836}}].

\bibitem{Shukla:16}
H.~{Shukla}, G.~{Mellema}, I.~T. {Iliev} and P.~R. {Shapiro}, \emph{{The
  effects of Lyman-limit systems on the evolution and observability of the
  epoch of reionization}},
  \href{http://dx.doi.org/10.1093/mnras/stw249}{\emph{\mnras} {\bf 458} (May,
  2016) 135--150}, [\href{http://arxiv.org/abs/1602.01144}{{\tt 1602.01144}}].

\bibitem{Bose:16b}
S.~{Bose}, C.~S. {Frenk}, H.~{Jun}, C.~G. {Lacey} and M.~R. {Lovell},
  \emph{{Reionisation in sterile neutrino cosmologies}}, {\emph{ArXiv e-prints}
  (May, 2016) }, [\href{http://arxiv.org/abs/1605.03179}{{\tt 1605.03179}}].

\bibitem{Wise:12}
J.~H. {Wise}, T.~{Abel}, M.~J. {Turk}, M.~L. {Norman} and B.~D. {Smith},
  \emph{{The birth of a galaxy - II. The role of radiation pressure}},
  \href{http://dx.doi.org/10.1111/j.1365-2966.2012.21809.x}{\emph{\mnras} {\bf
  427} (Nov., 2012) 311--326}, [\href{http://arxiv.org/abs/1206.1043}{{\tt
  1206.1043}}].

\bibitem{Johnson:12}
J.~L. {Johnson}, C.~{Dalla Vecchia} and S.~{Khochfar}, \emph{{The First Billion
  Years project: the impact of stellar radiation on the co-evolution of
  Populations II and III}},
  \href{http://dx.doi.org/10.1093/mnras/sts011}{\emph{\mnras} {\bf 428} (Jan.,
  2013) 1857--1872}, [\href{http://arxiv.org/abs/1206.5824}{{\tt 1206.5824}}].

\bibitem{Paardekooper:12}
J.-P. {Paardekooper}, S.~{Khochfar} and C.~{Dalla Vecchia}, \emph{{The First
  Billion Years project: proto-galaxies reionizing the Universe}},
  \href{http://dx.doi.org/10.1093/mnrasl/sls032}{\emph{\mnras} {\bf 429} (Feb.,
  2013) L94--L98}, [\href{http://arxiv.org/abs/1211.1670}{{\tt 1211.1670}}].

\bibitem{Yajima:14}
H.~{Yajima} and S.~{Khochfar}, \emph{{Can the 21-cm signal probe Population III
  and II star formation?}},
  \href{http://dx.doi.org/10.1093/mnras/stu2687}{\emph{\mnras} {\bf 448} (Mar.,
  2015) 654--665}, [\href{http://arxiv.org/abs/1405.7385}{{\tt 1405.7385}}].

\bibitem{Giallongo:15}
E.~{Giallongo}, A.~{Grazian}, F.~{Fiore}, A.~{Fontana}, L.~{Pentericci},
  E.~{Vanzella} et~al., \emph{{Faint AGNs at z $\geq$ 4 in the CANDELS GOODS-S
  field: looking for contributors to the reionization of the Universe}},
  \href{http://dx.doi.org/10.1051/0004-6361/201425334}{\emph{\aap} {\bf 578}
  (June, 2015) A83}, [\href{http://arxiv.org/abs/1502.02562}{{\tt
  1502.02562}}].

\bibitem{Finlator:16}
K.~{Finlator}, B.~D. {Oppenheimer}, R.~{Dav{\'e}}, E.~{Zackrisson},
  R.~{Thompson} and S.~{Huang}, \emph{{The soft, fluctuating UVB at z $\sim$ 6
  as traced by C IV, Si IV, and C II}},
  \href{http://dx.doi.org/10.1093/mnras/stw805}{\emph{\mnras} {\bf 459} (July,
  2016) 2299--2310}.

\bibitem{McQuinn:12}
M.~{McQuinn}, \emph{{Constraints on X-ray emissions from the reionization
  era}},
  \href{http://dx.doi.org/10.1111/j.1365-2966.2012.21792.x}{\emph{\mnras} {\bf
  426} (Oct., 2012) 1349--1360}, [\href{http://arxiv.org/abs/1206.1335}{{\tt
  1206.1335}}].

\bibitem{Dolgov:00}
A.~D. {Dolgov} and S.~H. {Hansen}, \emph{{Massive sterile neutrinos as warm
  dark matter}},
  \href{http://dx.doi.org/10.1016/S0927-6505(01)00115-3}{\emph{Astroparticle
  Physics} {\bf 16} (Jan., 2002) 339--344},
  [\href{http://arxiv.org/abs/hep-ph/0009083}{{\tt hep-ph/0009083}}].

\bibitem{Abazajian:01a}
K.~{Abazajian}, G.~M. {Fuller} and W.~H. {Tucker}, \emph{{Direct Detection of
  Warm Dark Matter in the X-Ray}},
  \href{http://dx.doi.org/10.1086/323867}{\emph{\apj} {\bf 562} (Dec., 2001)
  593--604}, [\href{http://arxiv.org/abs/astro-ph/0106002}{{\tt
  astro-ph/0106002}}].

\bibitem{Hansen:03}
S.~H. {Hansen} and Z.~{Haiman}, \emph{{Do We Need Stars to Reionize the
  Universe at High Redshifts? Early Reionization by Decaying Heavy Sterile
  Neutrinos}}, \href{http://dx.doi.org/10.1086/379636}{\emph{\apj} {\bf 600}
  (Jan., 2004) 26--31}, [\href{http://arxiv.org/abs/astro-ph/0305126}{{\tt
  astro-ph/0305126}}].

\bibitem{Pierpaoli:03}
E.~{Pierpaoli}, \emph{{Decaying Particles and the Reionization History of the
  Universe}},
  \href{http://dx.doi.org/10.1103/PhysRevLett.92.031301}{\emph{Physical Review
  Letters} {\bf 92} (Jan., 2004) 031301},
  [\href{http://arxiv.org/abs/astro-ph/0310375}{{\tt astro-ph/0310375}}].

\bibitem{Liu:16}
H.~{Liu}, T.~R. {Slatyer} and J.~{Zavala}, \emph{{The Darkest Hour Before Dawn:
  Contributions to Cosmic Reionisation from Dark Matter Annihilation and
  Decay}}, {\emph{ArXiv e-prints} (Apr., 2016) },
  [\href{http://arxiv.org/abs/1604.02457}{{\tt 1604.02457}}].

\bibitem{Oldengott:16}
I.~M. {Oldengott}, D.~{Boriero} and D.~J. {Schwarz}, \emph{{Reionization and
  dark matter decay}}, {\emph{ArXiv e-prints} (May, 2016) },
  [\href{http://arxiv.org/abs/1605.03928}{{\tt 1605.03928}}].

\bibitem{Adams:98}
J.~A. {Adams}, S.~{Sarkar} and D.~W. {Sciama}, \emph{{Cosmic microwave
  background anisotropy in the decaying neutrino cosmology}},
  \href{http://dx.doi.org/10.1046/j.1365-8711.1998.02017.x}{\emph{\mnras} {\bf
  301} (Nov., 1998) 210--214},
  [\href{http://arxiv.org/abs/astro-ph/9805108}{{\tt astro-ph/9805108}}].

\bibitem{Chen:03}
X.~{Chen} and M.~{Kamionkowski}, \emph{{Particle decays during the cosmic dark
  ages}}, \href{http://dx.doi.org/10.1103/PhysRevD.70.043502}{\emph{\prd} {\bf
  70} (Aug., 2004) 043502--+},
  [\href{http://arxiv.org/abs/arXiv:astro-ph/0310473}{{\tt
  arXiv:astro-ph/0310473}}].

\bibitem{Padmanabhan:05}
N.~{Padmanabhan} and D.~P. {Finkbeiner}, \emph{{Detecting dark matter
  annihilation with CMB polarization: Signatures and experimental prospects}},
  \href{http://dx.doi.org/10.1103/PhysRevD.72.023508}{\emph{\prd} {\bf 72}
  (July, 2005) 023508}, [\href{http://arxiv.org/abs/astro-ph/0503486}{{\tt
  astro-ph/0503486}}].

\bibitem{Biermann:06}
P.~L. {Biermann} and A.~{Kusenko}, \emph{{Relic keV Sterile Neutrinos and
  Reionization}},
  \href{http://dx.doi.org/10.1103/PhysRevLett.96.091301}{\emph{Physical Review
  Letters} {\bf 96} (Mar., 2006) 091301},
  [\href{http://arxiv.org/abs/astro-ph/0601004}{{\tt astro-ph/0601004}}].

\bibitem{Ripamonti:06b}
E.~{Ripamonti}, M.~{Mapelli} and A.~{Ferrara}, \emph{{The impact of dark matter
  decays and annihilations on the formation of the first structures}},
  \href{http://dx.doi.org/10.1111/j.1365-2966.2006.11402.x}{\emph{\mnras} {\bf
  375} (Mar., 2007) 1399--1408},
  [\href{http://arxiv.org/abs/astro-ph/0606483}{{\tt astro-ph/0606483}}].

\bibitem{Sitwell:13}
M.~{Sitwell}, A.~{Mesinger}, Y.-Z. {Ma} and K.~{Sigurdson}, \emph{{The imprint
  of warm dark matter on the cosmological 21-cm signal}},
  \href{http://dx.doi.org/10.1093/mnras/stt2392}{\emph{\mnras} {\bf 438} (Mar.,
  2014) 2664--2671}, [\href{http://arxiv.org/abs/1310.0029}{{\tt 1310.0029}}].

\bibitem{Furlanetto:15a}
S.~R. {Furlanetto}, \emph{{The 21-cm Line as a Probe of Reionization}},  in
  \emph{Astrophysics and Space Science Library} (A.~{Mesinger}, ed.), vol.~423
  of \emph{Astrophysics and Space Science Library}, p.~247, 2016.
\newblock \href{http://arxiv.org/abs/1511.01131}{{\tt 1511.01131}}.
\newblock \href{http://dx.doi.org/10.1007/978-3-319-21957-8_9}{DOI}.

\bibitem{Liu:15}
A.~{Liu}, J.~R. {Pritchard}, R.~{Allison}, A.~R. {Parsons}, U.~{Seljak} and
  B.~D. {Sherwin}, \emph{{Eliminating the optical depth nuisance from the CMB
  with 21 cm cosmology}},
  \href{http://dx.doi.org/10.1103/PhysRevD.93.043013}{\emph{\prd} {\bf 93}
  (Feb., 2016) 043013}, [\href{http://arxiv.org/abs/1509.08463}{{\tt
  1509.08463}}].

\bibitem{Fialkov:16}
A.~{Fialkov} and A.~{Loeb}, \emph{{Precise Measurement of the Reionization
  Optical Depth from The Global 21-cm Signal Accounting for Cosmic Heating}},
  {\emph{ArXiv e-prints} (Jan., 2016) },
  [\href{http://arxiv.org/abs/1601.03058}{{\tt 1601.03058}}].

\bibitem{Park:13}
H.~{Park}, P.~R. {Shapiro}, E.~{Komatsu}, I.~T. {Iliev}, K.~{Ahn} and
  G.~{Mellema}, \emph{{The Kinetic Sunyaev-Zel'dovich Effect as a Probe of the
  Physics of Cosmic Reionization: The Effect of Self-regulated Reionization}},
  \href{http://dx.doi.org/10.1088/0004-637X/769/2/93}{\emph{\apj} {\bf 769}
  (June, 2013) 93}, [\href{http://arxiv.org/abs/1301.3607}{{\tt 1301.3607}}].

\bibitem{Calabrese:14}
E.~{Calabrese}, R.~{Hlo{\v z}ek}, N.~{Battaglia}, J.~R. {Bond}, F.~{de
  Bernardis}, M.~J. {Devlin} et~al., \emph{{Precision epoch of reionization
  studies with next-generation CMB experiments}},
  \href{http://dx.doi.org/10.1088/1475-7516/2014/08/010}{\emph{\jcap} {\bf 8}
  (Aug., 2014) 010}, [\href{http://arxiv.org/abs/1406.4794}{{\tt 1406.4794}}].

\bibitem{Alvarez:15}
M.~A. {Alvarez}, \emph{{The Kinetic Sunyaev-Zel'dovich Effect from
  Reionization: Simulated Full Sky Maps at Arcminute Resolution}}, {\emph{ArXiv
  e-prints} (Nov., 2015) }, [\href{http://arxiv.org/abs/1511.02846}{{\tt
  1511.02846}}].

\bibitem{George:14}
E.~M. {George}, C.~L. {Reichardt}, K.~A. {Aird}, B.~A. {Benson}, L.~E. {Bleem},
  J.~E. {Carlstrom} et~al., \emph{{A Measurement of Secondary Cosmic Microwave
  Background Anisotropies from the 2500 Square-degree SPT-SZ Survey}},
  \href{http://dx.doi.org/10.1088/0004-637X/799/2/177}{\emph{\apj} {\bf 799}
  (Feb., 2015) 177}, [\href{http://arxiv.org/abs/1408.3161}{{\tt 1408.3161}}].

\bibitem{Paranjape:14}
A.~{Paranjape} and T.~R. {Choudhury}, \emph{{An improved model of H II bubbles
  during the epoch of reionization}},
  \href{http://dx.doi.org/10.1093/mnras/stu911}{\emph{\mnras} {\bf 442} (Aug.,
  2014) 1470--1482}, [\href{http://arxiv.org/abs/1401.7994}{{\tt 1401.7994}}].

\bibitem{Xu:13}
Y.~{Xu}, B.~{Yue}, M.~{Su}, Z.~{Fan} and X.~{Chen}, \emph{{An Analytical Model
  of the Large Neutral Regions during the Late Stage of Reionization}},
  \href{http://dx.doi.org/10.1088/0004-637X/781/2/97}{\emph{\apj} {\bf 781}
  (Feb., 2014) 97}, [\href{http://arxiv.org/abs/1312.4960}{{\tt 1312.4960}}].

\bibitem{Cole:00}
S.~{Cole}, C.~G. {Lacey}, C.~M. {Baugh} and C.~S. {Frenk}, \emph{{Hierarchical
  galaxy formation}},
  \href{http://dx.doi.org/10.1046/j.1365-8711.2000.03879.x}{\emph{\mnras} {\bf
  319} (Nov., 2000) 168--204},
  [\href{http://arxiv.org/abs/astro-ph/0007281}{{\tt astro-ph/0007281}}].

\bibitem{Lacey:15}
C.~G. {Lacey}, C.~M. {Baugh}, C.~S. {Frenk}, A.~J. {Benson}, R.~G. {Bower},
  S.~{Cole} et~al., \emph{{A unified multi-wavelength model of galaxy
  formation}}, {\emph{ArXiv e-prints} (Sept., 2015) },
  [\href{http://arxiv.org/abs/1509.08473}{{\tt 1509.08473}}].

\bibitem{Gonzalez-Perez:13}
V.~{Gonzalez-Perez}, C.~G. {Lacey}, C.~M. {Baugh}, C.~D.~P. {Lagos},
  J.~{Helly}, D.~J.~R. {Campbell} et~al., \emph{{How sensitive are predicted
  galaxy luminosities to the choice of stellar population synthesis model?}},
  \href{http://dx.doi.org/10.1093/mnras/stt2410}{\emph{\mnras} {\bf 439} (Mar.,
  2014) 264--283}, [\href{http://arxiv.org/abs/1309.7057}{{\tt 1309.7057}}].

\bibitem{Kennedy:13}
R.~{Kennedy}, C.~{Frenk}, S.~{Cole} and A.~{Benson}, \emph{{Constraining the
  warm dark matter particle mass with Milky Way satellites}},
  \href{http://dx.doi.org/10.1093/mnras/stu719}{\emph{\mnras} {\bf 442} (Aug.,
  2014) 2487--2495}, [\href{http://arxiv.org/abs/1310.7739}{{\tt 1310.7739}}].

\bibitem{Kaurov:15b}
A.~A. {Kaurov}, \emph{{On improving analytical models of cosmic reionization
  for matching numerical simulation}}, {\emph{ArXiv e-prints} (Dec., 2015) },
  [\href{http://arxiv.org/abs/1512.01312}{{\tt 1512.01312}}].

\bibitem{Sarkar:15}
A.~{Sarkar}, R.~{Mondal}, S.~{Das}, S.~K. {Sethi}, S.~{Bharadwaj} and D.~J.~E.
  {Marsh}, \emph{{The effects of the small-scale DM power on the cosmological
  neutral hydrogen ($\backslash$HI) distribution at high redshifts}},
  {\emph{ArXiv e-prints} (Dec., 2015) },
  [\href{http://arxiv.org/abs/1512.03325}{{\tt 1512.03325}}].

\bibitem{Greig:16}
B.~{Greig} and A.~{Mesinger}, \emph{{The Global History of Reionisation}},
  {\emph{ArXiv e-prints} (May, 2016) },
  [\href{http://arxiv.org/abs/1605.05374}{{\tt 1605.05374}}].

\bibitem{Finlator:14}
K.~{Finlator}, R.~{Thompson}, S.~{Huang}, R.~{Dav{\'e}}, E.~{Zackrisson} and
  B.~D. {Oppenheimer}, \emph{{The reionization of carbon}},
  \href{http://dx.doi.org/10.1093/mnras/stu2668}{\emph{\mnras} {\bf 447} (Mar.,
  2015) 2526--2539}, [\href{http://arxiv.org/abs/1412.4810}{{\tt 1412.4810}}].

\bibitem{Ocvirk:15}
P.~{Ocvirk}, N.~{Gillet}, P.~R. {Shapiro}, D.~{Aubert}, I.~T. {Iliev},
  R.~{Teyssier} et~al., \emph{{Cosmic Dawn (CoDa): the First
  Radiation-Hydrodynamics Simulation of Reionization and Galaxy Formation in
  the Local Universe}}, {\emph{ArXiv e-prints} (Oct., 2015) },
  [\href{http://arxiv.org/abs/1511.00011}{{\tt 1511.00011}}].

\end{thebibliography}\endgroup

\end{document}